%% file: MonteCarlo.tex
\documentclass[12pt,a4paper]{article}
\pdfoutput=1
\usepackage[latin1]{inputenc}
\usepackage{amsmath}
\usepackage{amsfonts}
\usepackage{amssymb}
\usepackage{makeidx}
\usepackage[T1]{fontenc}
\usepackage{bm}
\usepackage{bbm}
\usepackage{latexsym}
\usepackage{enumerate}
\usepackage{color}
\usepackage{ulem}
\usepackage{graphicx}
\usepackage{epsfig}

\input{macro}

\begin{document}


\title{Spatio-temporal spike trains analysis for large scale networks using maximum entropy principle and Monte-Carlo method}

\author
{
Hassan Nasser \thanks{NeuroMathComp, INRIA, 2004 Route des Lucioles, 06902 Sophia-Antipolis, France.},
Olivier Marre \thanks{Institut de la Vision,17 rue Moreau, 75012 Paris, France. },
Bruno Cessac  \thanks{NeuroMathComp, INRIA, 2004 Route des Lucioles, 06902 Sophia-Antipolis, France.},
}
\maketitle

\begin{abstract}

Understanding the dynamics of neural networks is a major challenge in experimental neuroscience. For that purpose, a modelling of the recorded activity that reproduces
the main statistics of the data is required. In a first part, we present a review on recent results dealing with  spike train statistics analysis using maximum entropy
models (MaxEnt). Most of these studies have been focusing on modelling synchronous spike patterns, leaving aside the temporal dynamics of the neural activity. However,
the maximum entropy principle can be generalized to the temporal case, leading to Markovian models where memory effects and time correlations in the dynamics are properly 
taken into account. In a second part, we present a new method based on Monte-Carlo sampling which is suited for the fitting of large-scale spatio-temporal MaxEnt models. 
The formalism and the tools presented here will be essential to fit MaxEnt spatio-temporal models to large neural ensembles. 

\end{abstract}

\input{Introduction}
\input{MaxEnt}
\input{Method}
\input{Tests}
\input{Conclusion}

\bigskip

\small{
\textbf{Acknowledgments}
We are grateful to G. Tkacik, T. Mora, S. Kraria, T. Vi\'eville, F. Hebert for helpful advices and help.
This work was supported by the INRIA, ERC-NERVI number 227747, KEOPS ANR-CONICYT and European Union Project $\#$ FP7-269921 (BrainScales) projects to B.C and H.N.and ANR OPTIMA to O.M.
Finally, we would like to thank the reviewers for helpful comments and remarks.
}
\bibliographystyle{plain}
\bibliography{odyssee,biblio}

\end{document}

%% file: macro.tex
\newcommand\seq[3]{{#1}_{#2}^{#3}}
\newcommand{\bd}{\begin{document}}
\newcommand{\ed}{\end{document}}
\newcommand{\beq}{\begin{equation}}
\newcommand{\eeq}{\end{equation}}
\newcommand{\bef}{\begin{figure}}
\newcommand{\enf}{\end{figure}}
\newcommand{\bea}{\begin{eqnarray}}
\newcommand{\eea}{\end{eqnarray}}
\newcommand{\baR}{\begin{array}}
\newcommand{\eaR}{\end{array}}
\newcommand{\bc}{\begin{center}}
\newcommand{\ec}{\end{center}}
\newcommand{\ben}{\begin{enumerate}}
\newcommand{\een}{\end{enumerate}}
\newcommand{\bit}{\begin{itemize}}
\newcommand{\eit}{\end{itemize}}
\newcommand{\su}{\section}
\newcommand{\ssu}{\subsection}
\newcommand{\sssu}{\subsubsection}
\newcommand\ssssu[1]{\textbf{#1}}
\newcommand{\nid}{\noindent}
\newcommand{\nnb}{\nonumber}
\newcommand\bloc[2]{{\omega}_{#1}^{#2}}
\newcommand\blocp[2]{{\bra{\sigma \, \omega}}_{#1}^{#2}}
\newcommand{\un}{\mathbbm{1}}
\newcommand{\setZ}{\mathbbm{Z}}
\newcommand{\setN}{\mathbbm{N}}
\newcommand{\setR}{\mathbbm{R}}
\newcommand{\pare}[1]{\left(\, #1 \, \right)}
\newcommand{\scal}[1]{\langle\, #1 \, \rangle}
\newcommand{\bra}[1]{\left[\, #1 \, \right]}
\newcommand{\brac}[2]{\left[\, #1 \, | \, #2 \right]}
\newcommand{\Set}[1]{\left\{\, #1 \, \right\}}
\newcommand{\Setc}[2]{\left\{\, #1 \, ; \, #2 \right\}}
\newcommand{\Prob}[1]{P\left[\, #1 \, \right]}
\newcommand{\PT}[2]{\pi^{(T)}_{#2}\left[\, #1 \, \right]}
\newcommand{\pT}{\pi^{(T)}}
\newcommand{\pTo}{\pi^{(T)}_\omega}
\newcommand{\Probex}[1]{P^{(\ast)}\left[\, #1 \, \right]}
\newcommand{\probc}[2]{P[#1 \, \left| \, #2 \right.]}
\newcommand{\Probc}[2]{P\bra{#1 \, \left| \, #2 \right.}}
\newcommand{\Probcex}[2]{P^{(\ast)}\bra{#1 \, \left| \, #2 \right.}}
\newcommand{\Probcp}[2]{P^{(ap)}\bra{#1 \, \left| \, #2 \right.}}
\newcommand{\abs}[1]{\left|\, #1 \, \right|}
\newcommand{\deq}{\stackrel {\rm def}{=}}
\newcommand{\sqsubsetneq}{\stackrel {\tiny\sqsubset}{\tiny \neq}}
\newcommand\moy[1]{\mu\bra{#1}}
\newcommand\cB{{\cal B}}
\newcommand\cC{{\cal C}}
\newcommand\cD{{\cal D}}
\newcommand\cE{{\cal E}}
\renewcommand\H{{\cH_\bbeta}}
\newcommand\mub{{\mu_\bbeta}}
\renewcommand{\sb}{s_\bbeta}
\newcommand\cG{{\cal G}}
\newcommand\cH{{\cal H}}
\newcommand\cK{{\cal K}}
\newcommand\cL{{\cal L}}
\newcommand\cM{{\cal M}}
\newcommand\cS{{\cal S}}
\newcommand\cP{{\cal P}}
\newcommand\cO{{\cal O}}
\newcommand\cU{{\cal U}}
\newcommand\cW{{\cal W}}
\newcommand\f{{\bm{f}}}
\newcommand\bF{{\bm{F}}}
\newcommand\bo{{\bm{o}}}
\newcommand\blambda{{\bm{\lambda}}}
\newcommand\bbeta{{\bm{\beta}}}
\newcommand\gk[1]{g_k\pare{#1}}
\newcommand\gLk{g_{L,k}}
\newcommand\akj[1]{\alpha_{kj}(#1)}
\newcommand\sif[1]{\seq{\omega}{#1}{D}}
\newcommand\Xk[1]{X_k({#1})}
\newcommand\Ik[1]{I_k({#1})}
\newcommand\Skj[1]{S_{kj}({#1})}
\newcommand\XkD{\Xk{\bloc{0}{D-1}}}
\newcommand\Xkr{X_k^r}
\newcommand\tLk{\tau_{L,k}}
\newcommand\moyc[2]{\mu\bra{#1 \, | \, #2}}
\newcommand\Vkdet[1]{V_k^{(det)}({#1})}
\newcommand\Vksyn[1]{V_k^{(syn)}({#1})}
\newcommand\Vkext[1]{V_k^{(ext)}({#1})}
\newcommand\Vknoise[1]{V_k^{(noise)}({#1})}
\newcommand\sk[1]{\sigma_k({#1})}
\newcommand\skr{\sigma_k^r\pare{\bloc{0}{D-1}}}
\newcommand{\ent}[1]{\left[\, #1 \, \right]}
\newcommand\vm[1]{var_m\left[#1 \right]}
\newcommand\eg[1]{\stackrel {\rm #1}{=}}
\newcommand\tk[1]{\tau_k\pare{#1}}
\newcommand\tko{\tau_k(t,\omega)}
\newcommand\cBm[1]{\cB^{-1}\pare{#1}}
\newtheorem{theorem}{Theorem}
\newcommand{\bth}{\begin{theorem}}
\newcommand{\enth}{\end{theorem}}
\newcommand\omD[1]{\seq{\bra{\omega^{(#1)}}}{0}{D-1}}
\newcommand\omz[1]{\omega^{(#1)}(D)}
\newcommand\om[1]{\omega^{(#1)}}
\newcommand\Om[1]{\Omega^{(#1)}}
\newcommand\skjq{s_{kj;q}}
\newcommand\ikq{i_{k;q}}
\newcommand\skjqs{p_{kj;q_s}}
\newcommand\ikqs{i_{k;q_s}}
\newcommand\ikqsu{i_{k;q_{s_u}}}
\newcommand\ikqu{i_{k;q_u}}
\newcommand\xkq{x_{k;q}}
\newcommand\skjqv{s_{kj;q_{v}}}
\newcommand\tjr{t_j^{(r)}(\omega)}

\newcommand\rpf{R}
\newcommand\rpfc[1]{R\pare{#1}}
\newcommand\lpf{L}
\newcommand\lpfc[1]{L\pare{#1}}
\newcommand{\zb}{\zeta_{\bbeta}}
\newcommand{\etab}{\eta_{\bbeta}}
\newcommand{\mex}{\mu^{(\ast)}}
\newcommand{\phex}{\phi^{(\ast)}}
\newcommand{\Phex}{\Phi^{(\ast)}}
\newcommand{\Hex}{\cH_{\bbeta'}^{(\ast)}}
\newcommand\p[1]{\cP\bra{#1}}
\newcommand\w[2]{w_{#1}^{(#2)}}
\newcommand\Gb{\cG_\bbeta}
\newcommand\blp[3]{\omega^{#3,(#1)}_{#2}}
\newcommand\blpb[3]{\overline{\omega^{#3,(#1)}_{#2}}}
\newcommand\moyex[1]{\mu^{(\ast)}\bra{#1}}
\newcommand\E[1]{{\cal E}_{#1}}

%% file: Introduction.tex
\su{Introduction}

The structure of the cortical activity, and its relevance to sensory stimuli or motor planning, have been subject to long standing debate. While some studies tend to
demonstrate that the majority of the information conveyed by neurons is contained in the mean firing rate \cite{shadlen-newsome:98}, other works have shown evidence for a role
of the higher order neural assemblies in neural coding (\cite{singer-gray:95}, \cite{vaadia-etal:95}, \cite{abeles:82}, \cite{louie-wilson:01} and \cite{harris-etal:02}).

Many single cell studies have reported an irregular spiking activity which seems to be very close to a Poisson process; concluding that the activity spans a 
very large state space. Several studies claim that some specific patterns, called ``cortical songs'', appear in a recurrent fashion (\cite{ikegay-etal:04}), but their
existence is controversial (\cite{mokeichev-etal:07} and \cite{luczak-etal:09}), suggesting that the size of the state space explored by the activity could be
smaller than expected. This point requires an accurate description of the neural activity of populations of neurons (\cite{softky-Koch:93}, \cite{oram-etal:99}, \cite{lampl-etal:99}, \cite{tsodyks-etal:99} and \cite{kenet-etal:03}).

These controversies partially originate from the fact that characterizing the statistics of the neural activity observed during the simultaneous recording of several
neurons is challenging, since the number of possible  patterns grows exponentially with the number of neurons. As a consequence, the probability of each pattern cannot be 
reliably measured by empirical averaging, and  an underlying model is necessary to reduce the number of variable to be estimated. To infer the whole
state of the neural network, some attempts have been done to build a hidden dynamical model which would underlie the cortical responses of several recorded
neurons. Most of the time, this approach has been used to characterize the activity of neurons during different types of behaviour. 
Among others, Shenoy and colleagues \cite{shenoy-matthew-etal:11}
used a dynamical system to model the activities of multiple neurons recorded in the motor areas. Most of the time, in this approach, the number of neurons largely 
exceeds the number of parameters. The assumed low dimension of the underlying  dynamical system  is often due to the low dimension of the behavioural context itself.
For example, in a task where a monkey is asked to make a choice between a small number of options (e.g. moving toward one target amongst several), one can 
expect that the features of the neural activity which  are relevant to this task can be described with a number of parameters which is comparable to the number of 
possible actions. 

For more complex tasks or stimuli, the dimension of these models may have to be increased. This would be especially critical in the case of sensory networks stimulated
with natural or complex stimuli. For this latter issue, a different strategy has been proposed by Schneidman et al \cite{schneidman-berry-etal:06} and Shlens et al 
\cite{shlens-field-etal:06,shlens-field-etal:09}. Their purpose was to describe the statistics
of the retinal activity in response to natural stimuli. They defined a set of values (mean firing rates, correlations...) that must be fitted, and then picked the
\textit{least structured} of the models that would satisfy these constraints. This approach, which will be described below, is based on maximum entropy models of the 
activity. It is interesting to point out that, while the previous approach aims at finding a useful representation of the activity with the lowest dimension, the maximum 
entropy approach picks the model with the highest dimension. \\

In this paper, we first describe the challenge of modelling the statistics
of the neural activity, and review the results that were obtained using maximum
entropy models. Many studies focused on modelling the synchronous patterns, putting aside the issue of modelling the temporal dynamics of the neural activity. We show why the extent of maximum entropy models to the
 temporal case raises specific issues, such as treating correctly memory and time correlations, and how they can be solved.
The corresponding section (section \ref{SMaxEnt}) reviews the maximum entropy approach, and focuses on applying it to general spatio-temporal constraints. We also
include a short discussion on other spatio-temporal approaches to spike train statistics such as the Generalized Linear Model \cite{brillinger:88, mccullagh-nelder:89,
 paninski:04,truccolo-eden-etal:05,pillow-paninski-etal:05,pillow-shlens-etal:08,pillow-ahmadian-etal:11,pillow-ahmadian-etal:11b}. In a second part (section \ref{SMethod}) 
 we present  a new method, based on Monte-Carlo
sampling, which is suited for the fitting of large scale spatio-temporal models.   The section \ref{STests} provides examples and numerical tests in order to show how far
 we can go with the Monte-Carlo method and its performance.


%


%% file: MaxEnt.tex
\section{The maximum entropy principle}\label{SMaxEnt}

In this section, we present the maximum entropy principle in a general setting.  We first give a set of notations and definitions, then present a brief history of this
principle in spike train analysis. Finally, we introduce a framework which allows the handling of general spatio-temporal constraints. 

\ssu{Notations and definitions}\label{Sdef}
\sssu{Spike trains}

We consider a network of $N$ neurons.  We assume there is a minimal time scale $\delta$, such that a neuron can fire at most one spike within a time window of size
$\delta$. To each neuron $k$ and discrete time $n$, we associate a spike variable: $\omega_k(n)=1$ if  neuron $k$ fires at time $n$, and $\omega_k(n)=0$ otherwise. 
%
The state of the entire network in
time bin $n$ is thus described by a vector  $\omega(n) \deq \bra{\omega_k(n)}_{k=1}^{N}$, called a \textit{spiking pattern}.

A \textit{spike block}, which describes the activity of the whole network between to moment of time $n_1$ and $n_2$, is a finite ordered list of such vectors, written:
$$\bloc{n_1}{n_2} = \Set{\omega(n)}_{\{n_1 \leq n \leq n_2\}},$$
The \textit{range} of a block is $n_2-n_1+1$, the number of time steps from $n_1$ to $n_2$. Here is an 
example of a spike block of range $5$ with $N=4$ neurons.

$$
\tiny{\pare{
\begin{array}{cccccc}
1 & 1 & 0 & 1 & 0\\
0 & 1 & 0 & 1 & 0\\
0 & 0 & 0 & 1 & 0\\
0 & 1 & 0 & 1 & 1\\
\end{array}
}}
$$

A \textit{spike train} or \textit{raster} is a spike block $\bloc{0}{T-1}$ from some initial time $0$ to some final time $T-1$. To simplify notation we simply write 
$\omega$ for a spike train. We note $\Omega = \Set{0,1}^{NT}$ the set of all possible spike trains.

\sssu{Observables}

We call \textit{observable} a function
:
\beq \label{DefObs}
\cO(\omega)=\prod_{u=1}^r \omega_{k_u}(n_u),
\eeq
i.e. a product of binary spike events where $k_u$ is a neuron 
index and $n_u$ a time index, with $u=1, \dots, r$, for some integer $r>0$.
 Typical choices of observables are
$\omega_{k_1}(n_1)$ which is $1$ if neuron $k_1$ fires at time $n_1$ and which is $0$ otherwise;
$\omega_{k_1}(n_1) \, \omega_{k_2}(n_2)$  which is $1$  if neuron $k_1$ fires at time $n_1$ and neuron $k_2$ fires at time $n_2$ and which is $0$ otherwise. Another
example is $\omega_{k_1}(n_1) \,(1-\omega_{k_2}(n_2))$ which is $1$ is neuron $k_1$ fires at time $n_1$ \textit{and neuron $k_2$ is silent at time $n_2$}. This 
example emphasizes that observables are able to consider events where some neurons are silent.

We say that an observable $\cO$ has \textit{range $R$} if it depends on $R$ consecutive spike patterns, e.g. $\cO(\omega)=\cO(\bloc{0}{R-1})$.  We consider here that 
observables do not depend explicitly on time (\textit{time-translation invariance of observables}).
As a consequence, for any time $n$, $\cO(\bloc{0}{R-1})=\cO(\bloc{n}{n+R-1})$ whenever $\bloc{0}{R-1}=\bloc{n}{n+R-1}$.

\sssu{Spike train statistics}\label{SStats}

It is common in the study of spike trains to attempt to detect some statistical regularity. Spike trains statistics is assumed
to be summarized by a hidden probability  $\mu$ 
characterizing the probability of \textit{spatio-temporal} spike patterns: $\mu$ is defined as soon
as the probability $\moy{\bloc{n_1}{n_2}}$ of any block $\bloc{n_1}{n_2}$ is known. We assume that $\mu$ is
time-translation invariant: for any time $n$,
$\moy{\bloc{0}{R-1}}=\moy{\bloc{n}{n+R-1}}$, whenever $\bloc{0}{R-1}=\bloc{n}{n+R-1}$.

Equivalently, $\mu$ allows the computation of the average of the observables. We note $\moy{\cO}$
the average of the observable $\cO$ under $\mu$. If $\cO(\omega)=\omega_{k_1}(n_1)$ then $\moy{\cO}$
is the firing rate of neuron $k_1$ (it does not depend on $n_1$ from the time-translation invariance hypothesis);
if $\cO=\omega_{k_1}(n_1) \, \omega_{k_2}(n_2)$, then $\moy{\cO}$ is the probability that neurons $k_1$ and $k_2$ fire over the time span $n_2-n_1$. 
Additionally,  $\moy{\omega_{k_1}(0) \, \omega_{k_2}(0)}
-\moy{\omega_{k_1}(0)}\moy{\omega_{k_2}(0)}$ represents the instantaneous pairwise correlation between the neurons $k_1$ and $k_2$.

There are several methods which allow the computation or estimation of $\mu$. In the following we shall assume that neural activity is described by a Markov process with memory 
depth $D$ and positive time-translation invariant transition probabilities
 $\Probc{\omega(D)}{\bloc{0}{D-1}}>0$. 
From the assumption $\Probc{\omega(D)}{\bloc{0}{D-1}}>0$, this chain has a unique invariant probability $\mu$ such that, for any $n>D$, and any block $\bloc{0}{n}$:
\beq\label{ProbBlocks}
\moy{\bloc{0}{n}} = \prod_{l=0}^{n-D} \Probc{\omega(D+l)}{\bloc{l}{D+l-1}} \moy{\bloc{0}{D-1}}.
\eeq
Therefore, knowing the transition probabilities (corresponding to blocks $\bloc{0}{D}$ of range $D+1$)
and $\mu$ (which can be determined as well from the transition probabilities as exposed in section \ref{SMarkov}), the probability of larger blocks can be 
computed. Equation (\ref{ProbBlocks}) makes explicit the role of memory in statistics of spike blocks, via the product of transition probabilities and
the probability of the initial block $\moy{\bloc{0}{D-1}}$.

On the opposite, if $D=0$, the probability to have the spike pattern $\omega(D)$ does not depend on the past activity of the network (memory-less case). In this case
 $\Probc{\omega(D+l)}{\bloc{l}{D+l-1}}$ becomes $\moy{\omega(l)}$  and the probability (\ref{ProbBlocks}) of a block becomes: 
\beq\label{ProbBlocksInd}
\moy{\bloc{0}{n}} = \prod_{l=0}^{n} \moy{\omega(l)}.
\eeq
Therefore, in the memory-less case, spikes occurring at different times are \underline{independent}.  

This emphasizes the deep difference between the case $D=0$ and the case $D>0$. 


\sssu{Empirical average}\label{SEmpAv}

Let us assume that we are given an experimental raster of length $T$, such that $\bloc{0}{T-1}$. The 
estimation of spikes statistics has to be done on this sample. In the context of the maximum entropy principle, where statistics is assumed to be time translation
 invariant, statistics of events is obtained via \textit{time-average}. The time-average or empirical average of an observable $\cO$ in a raster $\omega$ of length $T$ is denoted by $\PT{\cO}{\omega}$.
%
%
For example, if $\cO=\omega_k(0)$ the time-average $\PT{\cO}{\omega}=
\frac{1}{T-1} \, \sum_{n=0}^{T-1} \omega_k(n)$
is the firing rate of neuron $k$, estimated on the experimental raster $\omega$.

The  empirical average  is a random variable, depending on the raster $\omega$,
 as well as on the time length of the sample and it has Gaussian fluctuations whose amplitude tends to 
$0$ as $T\to +\infty$ like $\frac{1}{\sqrt{T}}$. This is the case, e.g. for the empirical averages obtained from several spike train acquired with several repetition.

\sssu{Complexity of the set of spike blocks.}
If one has $N$ neurons and wants to consider spike block events within $R$ time steps, one has $2^{NR}$ possible states. For a reasonable Multi Electrodes Array (MEA) sample, $N=100$, $R=3$ 
(for a time lag of $30$ms with a $10$ms binning), this gives $2^{300} \sim 4 \times 10^{180}$, which is quite a bit more than the expected number of particles in the (visible)
 universe. Taking into account the huge number of states in the set of blocks, it is clear that any method requiring the extensive description of the state space will fail  
as $NR$ grows. Additionally, while the accessible state space is huge, the \textit{observed} state space (e.g.
 in an experimental raster) is rather small. For example, in a MEA raster for a retina experiment, the sample size $T$ is about $10^6-10^7$, which is quite a bit less 
than $2^{NR}$.
As a matter of fact, any reasonable estimation method must take this small-sample constraint into account. As we show in the next section, the maximum entropy 
principle and the related notion of Gibbs distributions allows us to take these aspects into account.

\ssu{The maximum entropy principle}


\sssu{Motivations} \label{SMot}

Following \ref{Sdef} the goal is to find a probability distribution $\mu$ such that:

\begin{itemize}

\item $\mu$ is inferred from an empirical raster $\omega$, by computing the empirical average of a set of ad hoc observables $\cO_k$, $k=1, \dots, K$. One asks
that the average of $\cO_k$ with respect to $\mu$ satisfies:
\beq\label{Constraints}
\moy{\cO_k}=\PT{\cO_k}{\omega}, \quad k=1, \dots, K.
\eeq
The mean of $\cO_k$, predicted by $\mu$ is equal to the 
mean computed on the experimental raster.
$\mu$ is called
a ``model'' in the sequel. The set of observables $\cO_k$ defines the model. 

\item $\mu$ has to be ``as simple as possible'', with the least structure and a minimum number of tunable parameters. In the maximum entropy paradigm \cite{jaynes:57}
 these issue 
are (partly) solved by seeking a probability distribution $\mu$ which maximizes the entropy under the constraints (\ref{Constraints}). The entropy is defined 
explicitly below (see eq. (\ref{entropie_spatial}),(\ref{Stat_Ent})).   

\item From the knowledge of $\mu$ one can compute the probability of arbitrary blocks (e.g. via eq. (\ref{ProbBlocks})) and the average of other observables than
 the $\cO_k$s. 

\end{itemize}

\nid \textbf{Remark.}

Assume that we want to select observables $\cO_k$ in the set of all possible observables with range $R$. For $N$ neurons there are  $2^{NR}$ possible choices.
 When $NR$ increases, the number of possible observables will quickly exceed the number of samples available in the recording. Including all of them in the model
 would overfit the data.
Therefore, one has to guide the choice of observables
by additional criteria.
We now review some of the criteria, which have been used by other authors.

\sssu{Spatial models}

In a seminal paper, Schneidman et al \cite{schneidman-berry-etal:06} aimed at unravelling the role of instantaneous pairwise correlations in retina 
spike trains. Although these correlations are weak, researchers investigated whether they play a more significant role in spike train statistics than firing rates.

Firing rates correspond to the average of observables of the form $\omega_i(0), i=1, \dots, N$ (the time index $0$ comes from the assumed time-translation invariance)
while instantaneous pairwise correlations correspond to averages of observables of the form $\omega_i(0)\omega_j(0), 1 \leq i < j \leq N$. Analysing the role of
pairwise correlations in spike train statistics, compared to firing rate, amounts therefore to comparing two models, defined by two different types of observables. 

Note that all of these observables correspond to spatial events occurring at the same time. They give no information on how the spike patterns at a given time  depend
on the past activity. This situation corresponds to a memory-less model ($D=0$ in section \ref{SStats}), where transition probabilities do not depend on the past. As a consequence
the sought probability $\mu$ weights blocks of range $1$, and the probability of blocks with larger range is given by (\ref{ProbBlocksInd}): spike patterns at 
successive time steps are \underline{independent} in spatial models.

In this case, the \textit{entropy} of $\mu$ is given by:
\beq\label{entropie_spatial}
S(\mu)=-\sum_{\omega(0)} \moy{\omega(0)} \log \moy{\omega(0)}.
\eeq
The natural $\log$ could be replaced by the logarithm in base $2$.\\

Now, the maximum entropy principle of Jaynes \cite{jaynes:57} corresponds to seeking a probability $\mu$ which maximizes $S(\mu)$ under the constraints 
(\ref{Constraints}). It can be shown (see section \ref{SMaxEntGen} for the general statement) that this maximization problem  
is equivalent to the following Lagrange problem: maximising the quantity $S(\mu)+\moy{\H}$ where $\H$, called a \textit{potential} is given by:

\beq\label{H}
\H \, = \, \sum_{k=1}^K \beta_k \cO_k.
\eeq

The $\beta_k$s are real numbers and free parameters. $\moy{\H}$ is the average of $\H$ with respect to $\mu$. Since $\H$ is a linear combination of observables we have
$\moy{\H}=\sum_{k=1}^K \beta_k \moy{\cO_k}$. If the $\cO_k$s have finite range and are $>-\infty$ and if the $\beta_k$s are finite then it can be shown (see section 
\ref{SMaxEntGen}) that there is only one probability $\mu$, depending on the $\beta_k$s, which solves the maximization problem. It is called a 
\textit{Gibbs distribution}.

In this context ($D=0$) it reads:
\beq\label{mu_spatial}
\moy{\omega(0)} \, = \, \frac{e^{\H(\omega(0))}}{Z_\bbeta},
 \eeq
where the normalization factor
\beq\label{Z}
Z_\bbeta=\sum_{\omega(0)} \,  e^{\H(\omega(0))},
\eeq
is the so-called \textit{partition function}.

To match (\ref{Constraints}) the parameters $\beta_k$ have to be tuned. This can be done thanks to the following property of $Z_\bbeta$:
\beq\label{dlogZ}
\moy{\cO_k}=\frac{\partial \log Z_\bbeta}{\partial \beta_k}.
\eeq
Thus the $\beta_k$s have to be tuned so that $\mu$
matches (\ref{Constraints}) as well as (\ref{dlogZ}).
It turns out that $\log Z_\bbeta$ is convex with respect to $\beta_k$s, so the problem has a unique solution.

Note that $\log Z_\bbeta$ does not only allow us to obtain the averages of the observables, it also allows to characterize fluctuations. If a raster is distributed according
 to the Gibbs distribution (\ref{mu_spatial}), then, as pointed out in section \ref{SEmpAv}, the empirical average of an observable has fluctuations. One can show that 
these fluctuations are Gaussian (Central limit theorem). The joint probability of $\PT{\cO_k}{\omega}$, $k=1, \dots, K$ is Gaussian, with mean $\moy{\cO_k}$  given by (\ref{dlogZ}) and 
covariance matrix $\frac{\Sigma}{T}$ where the matrix $\Sigma$ has entries:
\beq\label{d2logZ}
 \Sigma_{kl}=
 \frac{\partial^2 \log Z_\bbeta}{\partial \beta_k \partial \beta_l}.
\eeq

Let us now discuss what this principle gives in the two cases considered by Schneidman et al.

\begin{enumerate}[(i)]

\item\textbf{Only firing rates are constrained.} 
Then: 
$$
\H(\omega(0)) = \sum_{k=1}^N \beta_k \omega_k(0).
$$
It can be shown that the corresponding
probability $\mu$ is:
$$
\moy{\omega(0)}=\prod_{k=1}^{N} 
\frac{e^{\beta_k \, \omega_k(0)}}{1+e^{\beta_k}}.
$$
Thus, the corresponding  statistical model is such that spikes emitted by distinct neurons at the same time are independent.  The parameter $\beta_k$ is directly related to the
firing rate $r_k$ since $r_k=\mu\bra{\omega_k(0)=1}= \frac{e^{\beta_k}}{1+e^{\beta_k}}$, so that we have:

$$
\moy{\bloc{0}{n}} = \prod_{l=0}^n \, \prod_{k=1}^{N} \, 
r_k^{\omega_k(l)} \, (1-r_k)^{1-\omega_k(l)},  
$$

the classical probability of coin tossing with independent probabilities (\textit{Bernoulli model}).
Thus, fixing only the rates as constraints, the maximum entropy principle leads to analyze  spike statistics as if each spike were thrown randomly and independently,
 as with coin tossing. This is the
``most random model'', which has the advantage of making as few hypothesis as possible. However, when only constrained with mean firing rates, the prediction of even 
small spike blocks in the retina was not successful. This was expected since this model assumes independence between neurons, an assumption that has been proven wrong
in earlier studies (e.g. \cite{puchalla-schneidman-etal:05}). 

\item\textbf{Firing rates and pairwise correlations are constrained.}
In the second model, Schneidman et al constrained the maximum entropy model with both mean firing rates and instantaneous pairwise correlations between neurons.
In this case, 
$$
\H(\omega(0)) = \sum_{k=1}^N \beta_k \, \omega_k(0) + 
\sum_{k,l=1}^N \beta_{kl} \, \omega_k(0) \, \omega_l(0).
$$

Here the potential can be identified with the Hamiltonian of a magnetic system with binary spins. It is thus often called ``Ising model'' in the  spike train analysis 
literature, although the original Ising model has constant and positive couplings \cite{georgii:88}. The corresponding statistical model is the least structured model
respecting these first and second order pairwise instantaneous constraints. The number of parameters is of the order\footnote{Most approaches assumes moreover that the 
pairwise coefficients are symmetric $\beta_{kl}=\beta_{lk}$ which divides the number of parameters by $2$.} of $N^2$, to be compared with the $2^N$ possible
 patterns. 
 

\end{enumerate}

Schneidman et al showed that the Ising model model successfully predicts spatial patterns, a result which was confirmed by \cite{shlens-field-etal:06} (see 
\cite{nirenberg-victor-etal:07} for a review).
Other works have used the same method and found also a good prediction in cortical structure in vitro \cite{tang-jackson-etal:08}, in the visual cortex in vivo 
\cite{yu-huang-etal:08}.
Later on, several authors considered higher order terms 
still corresponding to $D=0$ (\cite{ohiorhenuan-mechler-etal:10,schneidman-berry-etal:06,tkacik-schneidman-etal:09,ganmor-segev-etal:11a}).
Note that these results have been obtained on relatively small subsets of neurons (usually groups of 10). An interesting challenge is to test how these results  hold 
for larger subsets of neurons, and if other constraints have to be added \cite{ganmor-segev-etal:11b}(Tkacik et al, in preparation). 

\sssu{One time step spatio-temporal models and detailed balance}

These models are only designed to predict the occurrence of ``spatial'' patterns, lying within one time bin.  The use of spatial observables naturally leads to a time
independence assumption where the probability of occurrence of a spatio-temporal pattern is given by the product  (\ref{ProbBlocksInd}). Tang et al. \cite{tang-jackson-etal:08} tried to 
predict the temporal statistics of the neural activity with such a model and showed that it does not give a faithful description of temporal statistics. The idea to 
consider spatio-temporal observables then naturally emerges with the problem of generalising the  probability eq. (\ref{mu_spatial}) to that case.

From the statistical mechanics point of view, a natural extension consists of considering the space of rasters $\Omega$ as a lattice where one dimension is "space" 
(neurons index) and the other is time. 
The idea is then to consider a potential still of the form (\ref{H}) but where the observables correspond to spatio-temporal events. We assume that $\H$ has range
 $R=D+1$, $0 \leq D < + \infty$.
The potential of a spike block $\bloc{0}{n}$, $n\geq D$ is:
\beq\label{H_chain}
\H\pare{\bloc{0}{n}}=\sum_{l=0}^{n-D} \H\pare{\bloc{l}{D+l}}
\eeq
On this basis, restricting to the case where $D=1$
(one time step memory depth)
Marre et al  have proposed in \cite{marre-boustani-etal:09}  to
construct a Markov chain, where  transition probabilities
$\Probc{\omega(l+1)}{\omega(l)}$
are proportional to $e^{\H(\bloc{l}{l+1})}$.
If $\mu$ is the invariant probability of that chain,
the application of (\ref{ProbBlocks}) leads
to probability of blocks $\moy{\bloc{0}{n}}$,
proportional to $e^{\H\pare{\bloc{0}{n}}}$:
the probability of a block is proportional to the exponential of its potential ("energy"). This approach is therefore quite natural from the statistical mechanics 
point of view.

The main problem, however, is "what is the proportionality coefficient ?" As shown in \cite{marre-boustani-etal:09}, the normalization of conditional probabilities 
does not reduce to the mere division by a constant partition function. This normalization factor is itself dependent on the past activity. 

To overcome this dependency, Marre et al assumed that the activity respected a detailed balance. In this particular case, it can be shown that the normalization 
factor becomes, again, a constant. But this is an important reduction that could have implications for the interpretation of the data: for example, with this simplification, it is not possible to give an account of assymetric cross-correlograms.

\ssu{General spatio-temporal models}\label{SMaxEntGen}

We now present the general formalism which allows to solve the variational problem "maximising entropy under spatio-temporal constraints". This approach is rigorous
 and the normalization problem is resolved without requiring additional assumptions such as detailed balance. At the end of this section, we briefly discuss other 
approaches considering spatio-temporal statistics and their relations to potentials of the form (\ref{H}). 

\sssu{Constructing the Markov Chain} \label{SMarkov}

In this section we show how one can generate a Markov chain
where transition probabilities are proportional to $e^{\H(\bloc{l}{l+D})}$, for a potential $\H$ corresponding to spatio-temporal events. We also
solve the normalization problem. This construction
is well known and is based on the so-called transfer matrix (see e.g. \cite{georgii:88} for a presentation in the context of statistical physics; 
\cite{parry-pollicott:90}  for a presentation in the context of ergodic theory and \cite{vasquez-marre-etal:12} for a presentation in the context of spike trains
 analysis).

This matrix is constructed as follows.
Consider two spike blocks $w_1,w_2$ of range $D\geq 1$. The transition
$w_1 \to w_2$ is \textit{legal} if $w_1$ has the form $\omega(0)\bloc{1}{D-1}$ and $w_2$ has the form $\bloc{1}{D-1}\omega(D)$. The vectors $\omega(0),\omega(D)$
 are arbitrary but the block  $\bloc{1}{D-1}$ is common.  Here is an example of a legal transition : 
$$
\tiny{w_1 =\left[
\begin{array}{ccc}
0&0&1\\
0&1&1\\
\end{array}
\right] 
};
\,
\tiny{w_2 =\left[
\begin{array}{ccc}
0&1&1\\
1&1&0\\
\end{array}
\right]
}
.$$
 Here is an example of a forbidden transition
$$
\tiny{w_1 =\left[
\begin{array}{ccc}
0&0&1\\
0&1&1\\
\end{array}
\right] 
};
\,
\tiny{w_2 =\left[
\begin{array}{ccc}
0&1&1\\
0&1&0\\
\end{array}
\right]
}
.$$

Any block $\bloc{0}{D}$ of range $R=D+1$ can be viewed as a legal transition from the block $w_1=\bloc{0}{D-1}$ to the block 
$w_2=\bloc{1}{D}$ and in this case we write $\bloc{0}{D} \sim w_1w_2$.  \\

The \textit{transfer matrix} $\cL$ is defined as:
\beq\label{transfermatrix}
\cL_{w_1,w_2}=   
\left\{
\begin{array}{lll}
 e^{\H(\bloc{0}{D})}
\quad &\mbox{if} \quad
w_1, w_2 &\mbox{is legal with } \bloc{0}{D} \sim w_1w_2   \\
0, \quad &\mbox{otherwise}.
\end{array}
\right. .
\eeq

From the matrix $\cL$ the transition matrix of a Markov chain can be constructed, as we now show. 
Since observables are assumed to be bounded from below, $\H(\bloc{0}{D})>-\infty$, thus $e^{\H(\bloc{0}{D})} > 0$ for each legal transition. As a consequence of the 
Perron-Frobenius theorem \cite{gantmacher:66,seneta:06}, $\cL$ has a unique real positive eigenvalue $s_\bbeta$, strictly larger than the modulus of the other 
eigenvalues (with a positive gap), and with right, $\rpf$, and left, $\lpf$, eigenvectors: $\cL\rpf=\sb\rpf, \, \lpf\cL=\sb \lpf$, or, equivalently\footnote{The 
right eigenvector $\rpf$ has $2^{ND}$ entries $R_{w}$ corresponding to blocks of range $D$. It obeys $\sum_{w_2} \cL_{w_1 w_2} R_{w_2}=\sb R_{w_1}$, where $w_1=\bloc{0}{D-1}$ 
and where the sum runs over blocks $w_2=\bloc{1}{D}$. Since $\cL_{w_1 w_2}$ is non zero only if the entries $w_1,w_2$ have the block $\bloc{1}{D-1}$ in common, and since the right hand side ($\sb R_{w_1}$) fixes the value of $w_1$, this sum holds
in fact on all possible values of $\omega(D)$. The notation $R_w$, although natural, does not make explicit the block involved. This is problematic when one wants to handle 
equations such as (\ref{P_large_blocks}). As a consequence, we prefer to use the notation $\rpfc{block}$ to make explicit this dependence. The same remark holds mutatis mutandis for the left eigenvector.}:
$$
\sum_{\omega(D) \in \Set{0,1}^N} e^{\H\pare{\bloc{0}{D}}}
  \rpfc{\bloc{1}{D}} = \sb \rpfc{\bloc{0}{D-1}};
$$

$$
\sum_{\omega(0) \in \Set{0,1}^N} \lpfc{\bloc{0}{D-1}} e^{\H\pare{\bloc{0}{D}}}
   = \sb \lpfc{\bloc{1}{D}}.
$$

These eigenvectors have strictly positive  entries $\rpfc{.}>0$, $\lpfc{.}>0$, functions of blocks of range $D$. They can be chosen so that the scalar product 
$\scal{L,R}=1$. We define:
\beq\label{Pres_vs_s}
\cP(\H)=\log \sb.
\eeq
called "topological pressure". We discuss the origin of this term and its properties in the section \ref{Sremarks}.

To define a Markov chain from the transfer matrix $\cL$ (eq. \ref{transfermatrix} ) we introduce the \textit{normalised potential}: 
\beq\label{Phi_norm}
\phi(\bloc{0}{D})=\H(\bloc{0}{D}) - \Gb(\bloc{0}{D})
\eeq
with:
\beq\label{G}
\Gb(\bloc{0}{D}) = \log \rpfc{\bloc{0}{D-1}} - \log \rpfc{\bloc{1}{D}}+\log s_\bbeta,
\eeq
and a family of transition probabilities:
\beq\label{Prob_cond}
\Probc{\omega(D)}{\bloc{0}{D-1}} \deq e^{\phi(\bloc{0}{D})}>0.
\eeq

These transition probabilities define a Markov chain which admits a unique invariant probability:
\beq\label{mu_inv}
\mu(\bloc{0}{D-1})=\rpfc{\bloc{0}{D-1}}\lpfc{\bloc{0}{D-1}}.
\eeq

From the general form of block probabilities (\ref{ProbBlocks})  the probability of blocks of depth $n \geq D$ is, in this case :
\beq\label{pCondphi}
\moy{\bloc{0}{n}}  =
e^{\sum_{l=0}^{n-D} \phi\pare{\bloc{l}{D+l}}} \moy{\bloc{0}{D-1}}.
\eeq
thus, from (\ref{mu_inv}),(\ref{Phi_norm}),(\ref{G}):
\beq\label{P_large_blocks}
\moy{\bloc{0}{n}} = \frac{e^{\H\pare{\bloc{0}{n}}}}{\sb^{n-D+1}} 
\rpfc{\bloc{n-D+1}{n}}\lpfc{\bloc{0}{D-1}},
\eeq
where $\H\pare{\bloc{0}{n}}$ is given by (\ref{H_chain}).

\sssu{Remarks}\label{Sremarks}

\begin{enumerate}

\item We have been able to compute the probability of any blocks $\bloc{0}{n}$. It is proportional to $e^{\H\pare{\bloc{0}{n}}}$ and the proportionality factor has been computed. 
In the general case of spatio-temporal events, it depends on $\bloc{0}{D-1}$ and $\bloc{n-D+1}{n}$. 

The same  arises in statistical mechanics when dealing with boundary conditions. The forms (\ref{pCondphi}), (\ref{P_large_blocks}),
remind Gibbs distributions on spin lattices, with
lattice translation-invariant probability distributions given specific boundary conditions.
Given a spin-potential of spatial-range $n$, the probability of a spin block depends upon the state of the spin block, as well as spins states
in a neighbourhood of that block. The conditional probability of this block
given a fixed neighbourhood is the exponential of the energy characterizing physical interactions, within the block, as well as interactions with the boundaries. 
In (\ref{pCondphi}), spins are replaced by spiking patterns;
space is replaced by time.
Spatial boundary conditions are here replaced
by the dependence upon $\bloc{0}{D-1}$ and $\bloc{n-D+1}{n}$. 

As a consequence, as soon as one is dealing with spatio-temporal events the normalization of conditional probabilities does not reduce to the mere division by:
\beq\label{Zn}
Z_n =  \sum_{\bloc{0}{n}} e^{\H\pare{\bloc{0}{n}}},
\eeq
as easily checked in (\ref{P_large_blocks}). 

\item The topological pressure obeys nevertheless:
%
%
%
$$
\cP(\H) = \lim_{n \to +\infty} \frac{1}{n} \log Z_n,
$$
and is analogue to a thermodynamic potential density (free energy, free enthalpy, pressure). This analogy is also clear in the variational principle (\ref{VarPrinc}) below.
 To our best knowledge the term "topological pressure" has its roots in the thermodynamic formalism of hyperbolic (chaotic) maps \cite{ruelle:78,parry-pollicott:90,beck-schloegl:95}. 
In this context,  this function can be computed as the grand potential of the 
grand canonical ensemble, as a cycle expansion over unstable periodic orbits. It is therefore equivalent to a pressure\footnote{The grand potential $\Phi$
obeys $\Phi=-PV$, where $P$ is the physical pressure and $V$ the volume. Therefore, the grand potential density is (minus) the pressure.}
depending on topological properties (periodic orbits).

\item In the case $D=0$ the Gibbs distribution reduces to (\ref{mu_spatial}).  One can indeed easily show that:

$$ \exp \Gb=\sb = \sum_{\omega(0)} \,  e^{\H(\omega(0))}=Z_\bbeta,
$$
where $Z_\bbeta$ is the partition function (\ref{Z}). Additionally, since spike patterns occurring at distinct time are independent in the $D=0$ case, $Z_n$ in (\ref{Zn}) can be written as $Z_n = Z_\bbeta^n$ so that $\cP(\H)=\log Z_\bbeta$.

\item In the general case of spatio-temporal constraints, the normalization requires the consideration of normalizing function $\Gb$ \textit{depending as well on the
 blocks $\bloc{0}{D}$}.
Thus, in addition to function $\H$ normalization introduces a second \textit{function} of spike blocks.  
This increases consequently the complexity of Gibbs potentials and Gibbs distributions compared to the spatial ($D=0$) case where $\Gb$ reduces to a constant. 

\end{enumerate}

\sssu{The maximum entropy principle}

We now show that the probability distribution defined this way solves the variational problem ``maximising entropy under constraints''.

We define  the \textit{entropy rate} (or Kolmogorov-Sinai entropy):
\beq\label{Stat_Ent}
h\bra{\mu} \, = \, - \, \limsup_{n \to \infty} \frac{1}{n+1} \, \sum_{\bloc{0}{n}} \, \moy{\bloc{0}{n}} \, \log \moy{\bloc{0}{n}},
\eeq
where the sum holds over all possible blocks $\bloc{0}{n}$. Note, that in the case of a Markov chain $h\bra{\mu}$ also reads \cite{cornfeld-fomin-sinai:82}:
\beq\label{Ent_Markov}
h\bra{\mu} \, = \, - 
 \sum_{\bloc{0}{D}}
 \, \moy{\bloc{0}{D}} \,
  \Probc{\omega(D)}{\bloc{0}{D-1}} \,
 \log 
 \Probc{\omega(D)}{\bloc{0}{D-1}}, 
 \eeq
whereas,  when $D=0$, $h\bra{\mu}$ reduces to the  definition (\ref{entropie_spatial}).\\

As a general result from ergodic theory
\cite{ruelle:78,keller:98,chazottes-keller:09} and mathematical statistical physics \cite{georgii:88}, there is a unique\footnote{The result is straightforward here 
since we consider bounded potentials with finite range.} probability distribution $\mu$ such that
 \cite{ruelle:78,keller:98,chazottes-keller:09}:  
\beq\label{VarPrinc}
\p{\H}=\sup_{\nu \in \cM_{inv}} \pare{h\bra{\nu} \, + \, \nu\bra{\H)}}=
h\bra{\mu} \, + \, \moy{\H},
\eeq
where $\p{\H}$ is given by (\ref{Pres_vs_s}). $\cM_{inv}$ is the set of all possible time-translation invariant probabilities on the set of rasters with $N$ neurons 
and $\nu\bra{\H}=
\sum_{\bloc{0}{D}} \H(\bloc{0}{D}) \, \nu(\bloc{0}{D})$ is the average value of
$\H$ with respect to the probability $\nu$.

Looking at the second equality, the variational principle  (\ref{VarPrinc}) selects, among all possible probabilities $\nu$,
a unique one realizing the supremum. This is exactly the invariant distribution of the Markov chain and is the sought  Gibbs distribution. It is clear from (\ref{VarPrinc}) 
that the topological pressure is the formal analogue to a thermodynamic potential density,
where $\H$ somewhat fixes the "ensemble": $\nu\bra{\H} = \sum_{k=1}^K \beta_k \nu\bra{\cO_k}$
plays the role of $\beta E$ (canonical ensemble), $\beta E - \mu N$ (grand canonical ensemble), $\dots$  in thermodynamics \cite{beck-schloegl:95}. 

\sssu{Inferring the parameters $\beta_k$}\label{SInfer}
The inverse problem of finding the values of $\beta_k$s from the observables average measured on the data is a hard problem with no exact analytical solution. However, in the context of spatial models with pairwise interactions the wisdom, coming from statistical physics and especially Ising model and spin-glasses, as well as from the Boltzmann machine learning community, can be used. As a consequence, in this context, several strategies were proposed. 
Ackley et al  \cite{ackley-etal:85} proposed a technique to estimate the parameters of a Boltzmann machine.
This technique is effective for small networks but it is time consuming. In practice, the time necessary to learn the parameters increases exponentially with the number of units. 
To speed up the parameters estimation, analytical approximations of the inverse problem have been proposed, which express the parameters  $\beta_k$  as a nonlinear function of the 
correlations of the activity (see for example \cite{tanaka:98}, \cite{roudi-nirenberg-etal:09} , \cite{sessak-monasson:09}, \cite{peterson-anderson:87}, \cite{roudi-tyrcha-etal:09},
\cite{ackley-etal:85}, \cite{higuchi-mezard:09}, \cite{kappen-rodriguez:98}).

These methods do not give an exact result, but are computationally fast. We do not pretend to review all of them  here, but we quote a  few  prominent examples.
In \cite{sessak-monasson:09}, Sessak and Monasson proposed a systematic small-correlation expansion to solve the inverse Ising problem. They were able to compute couplings up to the third order in the correlations for generic magnetizations,
and to the seventh order in the case of zero magnetizations. Their resulting expansion outperforms existing algorithms on the Sherrington-Kirkpatrick spin-glass model \cite{sherrington-kirkpatrick:75}.

Based on a high-field expansion of the Ising thermodynamic potential, Cocco et al \cite{cocco-leibler-etal:09} designed an algorithm to calculate the parameters in a time polynomial with N, 
where the couplings are expressed as a weighted sum over the power of the correlations. They did not obtained a closed analytical expression, but their algorithm could run
in a time that was polynomial in the number of neurons. 

Other methods, based 
on Thouless-Anderson-Palmer equations  \cite{thouless-anderson-etal:77} and linear response 
\cite{kappen-rodriguez:98}, or information geometry \cite{tanaka:98}, initially proposed in the field of spin-glasses, have been adapted and applied to spike train analysis (see e.g. the work done by 
 Roudi and collaborators \cite{roudi-hertz:11}).

The success of these approximations depends on the dataset, and there is no a priori guarantee about their efficiency at finding the right values of the parameters. However,
by getting closer to the correct solution, they can potentially speed up the convergence of the learning by starting with a seed much closer to the real solution than if taking
a random starting point.

Note also that all the techniques mentioned above have been designed for the case where there is no temporal interaction (except \cite{cocco-leibler-etal:09,roudi-hertz:11} which are discussed in the section \ref{SOther}). 
Now, we explain how the parameters estimation can be done in the spatio-temporal models. \\


In the general case parameters $\beta_k$s can be determined thanks to the following properties. 

\begin{itemize}
\item $\p{\H}$ is a log generating function of  cumulants. First:

\beq\label{AvOk}
\frac{\partial \p{\H}}{\partial \beta_k} =
\mu\bra{\cO_k}.
\eeq
This is an extension of (\ref{dlogZ}) to the time-dependent case.

\item Second:
\beq\label{d2Pbeta}
\frac{\partial^2 \p{\H}}{\partial \beta_k \partial \beta_l} = \frac{\partial \moy{\cO_k}}{\partial \beta_l} =
\sum_{n=0}^{+\infty} C_{\cO_k \cO_l}(n),
\eeq
where $C_{\cO_k\,\cO_l}(n)$
%
%
is  the correlation function between the two observables $\cO_k$ and $\cO_l$ at time $n$.
Note that correlation functions decay exponentially fast whenever $\H$ has finite range. So that $\sum_{n=0}^{+\infty} C_{\cO_k\,\cO_l}(n) < +\infty$.

Eq. (\ref{d2Pbeta}) characterizes the variation in the average value of
$\cO_k$ when varying $\beta_l$ (linear response).
The corresponding matrix is a susceptibility matrix. It controls the  Gaussian fluctuations of observables around their mean (central limit theorem)
 \cite{ruelle:78,parry-pollicott:90,chazottes-keller:09}. 
This is the generalization of (\ref{d2logZ}) to the time dependent case. As a particular case, the fluctuations of the empirical average $\PT{\cO_k}{\omega}$ of
 $\cO_k$ around its mean $\moy{\cO_k}$
are Gaussian with a mean-square deviation $\frac{\sqrt{\moy{\cO_k}(1-\moy{\cO_k}}}{\sqrt{T}}$.

It is clear that the structure of the linear response in the case of spatio-temporal constraints is quite a bit more complex than the case $D=0$ (see eq. (\ref{d2logZ})).
Actually, for $D=0$, all correlations $C_{\cO_k\,\cO_l}(n)$ vanish for $n>0$ (distinct times are independent).

\item 
$\cP(\H)$ is a convex function of $\bbeta$.  As a consequence, if there is a set of $\bbeta$ value,
$\bbeta^\ast$, such that .
\beq\label{Gradient}
\frac{\partial \p{\H}}{\partial \beta^*_k} =
\moy{\cO_k} = C_k,
\eeq
then this set is unique. Thus, the solution of the variational problem (\ref{VarPrinc}) is unique.
\end{itemize}

 Basically, eq. (\ref{AvOk}), (\ref{d2Pbeta}), \ref{Gradient}, tell us that techniques based on free energy expansion in spatial models can be extended as well to spatio-temporal cases, where the free energy is replaced by the topological pressure. Obviously, estimating (not to speak of computing) the topological pressure can be a formidable task. Although, the transfer matrix technique allows the computation of the topological pressure, the use of this method for large $N$ is hopeless (see section \ref{SLimitsTransfer}). However,
 techniques based on periodic orbit expansion and zeta functions,
 could be useful \cite{parry-pollicott:90}. Additionally, cumulant expansions of the pressure, eq. (\ref{AvOk}) and (\ref{d2Pbeta}) corresponding to the two first orders, suggest that extension of methods based on free energy expansion could be used. In addition to the works quoted above, we also think of
constraint satisfaction problems
by M\'ezard and Mora \cite{mezard-mora:09} and approaches based on Bethe free energy \cite{welling-teh:11}. Finally, as we checked, the properties of spatio-temporal Gibbs distributions allows to extend the parameters estimation methods developed for the spatial case in \cite{dudik-phillips-etal:04,broderick-dudik-etal:07} to spatio-temporal distributions (to be published).

\sssu{Other spatio-temporal models}\label{SOther}

Here we shortly review alternative spatio-temporal models. We essentially refer to approaches attempting to construct a Markov chain and related invariant probability
 by proposing a specific form for the transition probabilities.

A prominent example is provided by the so-called Linear-Nonlinear (LN) models and Generalized Linear Models (GLM)
 \cite{brillinger:88, mccullagh-nelder:89, paninski:04,truccolo-eden-etal:05,pillow-paninski-etal:05,pillow-shlens-etal:08,pillow-ahmadian-etal:11,pillow-ahmadian-etal:11b}.
 Shortly, the idea is to model spike statistics by a point process where the instantaneous firing rate of a neuron 
is a nonlinear function of the past network activity including feedbacks and interaction between neurons \cite{simoncelli-paninski-etal:04}.
This model has been applied in a wide variety of experimental
settings \cite{brillinger:92,chichilnisky:01,theunissen-david-etal:01,brown-barbieri-etal:03,paninski-fellows-etal:04,truccolo-eden-etal:05,
pillow-shlens-etal:08}.
Typically, referring e.g. to \cite{pillow-ahmadian-etal:11}, the rate $r_i$ has the form (adapting to our notations): 

\beq\label{RateGLM}
r_i = f\bra{
b_i + K_i . x + \sum_j
H_{ij}}
\eeq
where the kernel $K_i$ represents the $i$-th cell's linear receptive field and $x$ an input. $H_{ij}$ characterizes the effect of spikes emitted in 
the past by pre-synaptic neuron  $j$ on post-synaptic neuron $i$. 
In this approach, neurons are assumed to be conditionally-independent given  the past. 
The probability to have a given spike-response to a stimulus, given the past activity of the network, reads as the product of firing rates (see e.g. eq. 2.4 in \cite{pillow-ahmadian-etal:11}).

In \cite{pillow-ahmadian-etal:11} the authors use several Monte Carlo approaches to learn the parameters of the model for a Bayesian decoding of the 
rasters. Comparing to the method presented in the previous sections, the main advantages of the GLM is: (i) The transition probability is known (postulated) from 
the beginning and does not require the heavy normalization imposed by potentials of the form (\ref{H}); (ii) The model parameters have a  
neurophysiological interpretation, and their number grows at most as a power law in the number of neurons, as opposite to (\ref{H}), 
where the parameters are delicate to interpret and whose number can become quite large, depending on the set of constraints.

Note however that a model of the form (\ref{RateGLM}) can be written as well in the form (\ref{H}): 
this is a straightforward consequence of the Hammersley-Clifford theorem \cite{hammersley-clifford:71}. The parameters $\beta_
k$ in (\ref{H}) are then 
nonlinear functions of the parameters in (\ref{RateGLM}) (see \cite{cessac:11a} for an example).  \\

The main drawback of this approach is the assumption of conditional independence between neurons: neurons are assumed independent at time $t$ when the past, which appears in the function $H_{ij}$ in (\ref{RateGLM}), is given, and the probability of a spiking pattern at time $t$ is the product of neurons firing rates. On the opposite,  the maximal entropy principle does not require this assumption. 

It is interesting to remark that the conditional independence assumption can be rigorously justified in conductance based Integrate and Fire models 
\cite{cessac:11a,cessac:11b} and the form of the function $f$ can be explicitly found (this a sigmoid function instead of an exponential as usually 
postulated in GLM). This result holds true if only chemical synapses are involved (this is also implicit 
in the kernel form $H_{ij}$ in (\ref{RateGLM}) \cite{ pillow-ahmadian-etal:11}), but conditional independence breaks down for example as soon as electric synapses (gap junctions) are involved: this can be mathematically shown in
 conductance based Integrate and Fire models \cite{cofre-cessac:12}. Note that in this case, a large part of correlations is due dynamical interactions between neurons: as a consequence they persist even if there is no shared input. 
 
Recently, Macke et al. \cite{macke-etal:11}  extended the GLM model to fix the lack of 
instantaneous correlations between neurons in the GLM. They added a common input function that has a linear temporal dynamics. However,
one of the disadvantages of this technique is that its likelihood is not unimodal, and thus that computationally expensive Expectation-Maximization
algorithms have to be used to fit parameters.

The GLM model is usually used to model both the stimulus-response dependence as well as the
interaction between neurons, while the MaxEnt models usually focus on the latter (but see \cite{tkacik-prentice-etal:10}).\\

To finish this subsection, we would like to quote two important works dealing with spatio-temporal events too.  First, In \cite{cocco-leibler-etal:09}
Cocco and co-workers consider retinal ganglion cells spiking activity
with a dual approach: on one hand they consider an Ising model (and higher order spatial terms) where they propose an inverse method based on a cluster
 expansion to find efficiently the coupling in Ising model from data; on the other hand, they consider the problem of finding the parameters (synaptic 
couplings) in a Integrate and Fire model with noise, from its spike trains. In the weak noise limit the conditional probability of a spiking pattern 
given the past is given by a least action principle. This probability is
a Gibbs distribution whose normalized potential is characterized by the action computed over an optimal path. This second approach allows the characterization 
of spatio temporal events. Especially it gives a very good fit of the cross-correlograms.

Second, in \cite{roudi-hertz:11}, the authors consider a one step memory Markov chain where the conditional probability has a time-dependent potential of 
Ising type. Adapting a Thouless-Anderson-Palmer \cite{thouless-anderson-etal:77} approach used formerly in the Sherrington-Kirkpatrick mean-field model of 
spin glasses \cite{sherrington-kirkpatrick:75} they propose an inversion algorithm to find the model-parameters. As in the GLM their model assumes conditional
independence given the past   (see eq. (1) in \cite{roudi-hertz:11}).

\ssu{Comparing models}

Solving equations (\ref{Gradient}) provides an optimal choice for the Gibbs distribution $\mu$, \textit{given the observables} $\cO_k$. However, changing the set of
observables provides distinct Gibbs distributions, which does not approximate the hidden probability with the same accuracy. We need here a way to quantify the 
``distance'' between the ``model'' (the Gibbs distribution fixed by the set of observables) and the exact, hidden, probability $\mex$. Here are several criteria 
of comparison.

\sssu{Kullback-Leibler divergence}

The Kullback-Leibler divergence between $\mu,\mex$ is given by:
\beq\label{dKL}
d(\mex,\mu)=\limsup_{n \to \infty} \frac{1}{n+1}\sum_{\bloc{0}{n}} 
\mex\bra{\bloc{0}{n}}
\log\bra{\frac{\mex\bra{\bloc{0}{n}}}{\mu\bra{\bloc{0}{n}}}},
\eeq
which provides some notion of asymmetric ``distance'' between $\mu$ and $\mex$. The KL divergence accounts for discrepancy between the predicted probability
$\mu\bra{\bloc{0}{n}}$ and the exact probability $\mex\bra{\bloc{0}{n}}$ for all blocks of  range $n$. 

This quantity is not numerically computable from (\ref{dKL}). However, for $\mu$ a Gibbs distribution and $\mex$ a time-translation invariant probability, the 
following holds:

$$
d_{KL} \pare{\mex,\mu}  = \p{\H} \, - \, \mex\bra{\H} \, - \, h(\mex).
$$

The topological pressure $\p{\H}$ is given by (\ref{Pres_vs_s}) while $\mex\bra{\H}$ is estimated by $\PT{\H}{\omega}
= \sum_{k=1}^K \beta_k \PT{\cO_k}{\omega}=\sum_{k=1}^K \beta_k C_k$.

\begin{figure}[h!]
\begin{center} 
\includegraphics[height=7cm,width=7cm]{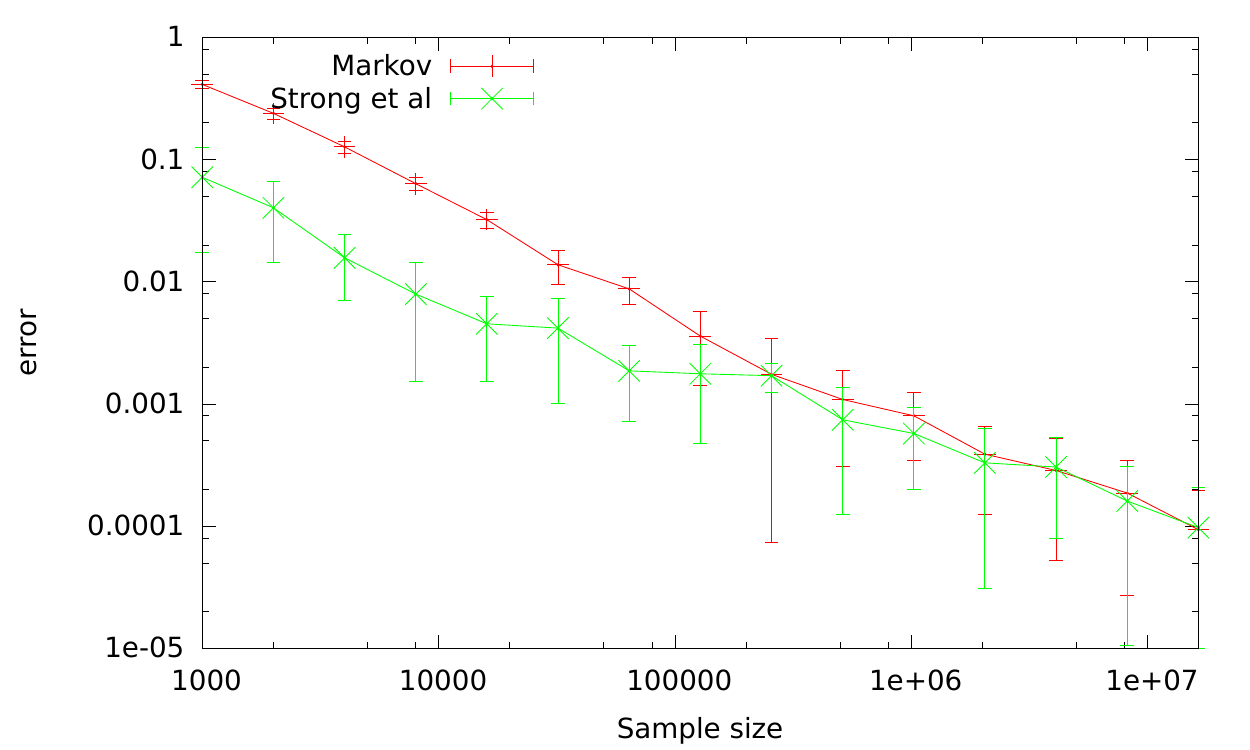} 
\caption{Difference between the exact probability and the approached entropy $h(\pTo)$, as a function of raster size $T$. The potential of test includes 
all the possible observables where weights are set as random values.}
\label{entropies}
\end{center} 
\end{figure}

Since $\mex$ is unknown, $ h(\mex)$ is unknown, and can only be estimated from data, i.e. one estimates the entropy of the empirical probability, $h(\pTo)$.
 There exist efficient methods for that. Note that the entropy of a Markov chain is readily given by eq. (\ref{Ent_Markov}), so the entropy $h(\pTo)$ is
 obtained by replacing the exact probability $P$ in eq. (\ref{Ent_Markov}), by the empirical probability $h(\pTo)$. As $T \to +\infty$, $h(\pTo)
 \to h(\mex)$, at exponential rate\footnote{The rate is given by the spectral gap of the transfer matrix: the difference between the largest eigenvalue (it is 
real and positive), and the modulus of the second largest eigenvalue (in modulus). \label{frate}}. For finite $T$ finite size corrections exist, see e.g. Strong et al. 
\cite{strong-koberle-etal:98}. In figure  \ref{entropies} is plotted an example. For a potential $\H$
with $N=5$ neurons and range $R=2$, containing all possible observables, we have plotted the difference between the exact probability (known from (\ref{Ent_Markov}) and the
 explicit form (\ref{Prob_cond}), (\ref{mu_inv}) of transition probabilities and invariant probability), and the approached entropy $h(\pTo)$ obtained
 by replacing  the exact probability $P$ by the empirical probability $h(\pTo)$, as a function of raster size $T$. We have also plotted the finite 
corrections method proposed by Strong et al. in \cite{strong-koberle-etal:98}.

Now, if one wants to compare how two Gibbs distributions $\mu_1,\mu_2$ approximate data, one compares the divergence $d_{KL} \pare{\mex,\mu_1}$, $d_{KL} 
\pare{\mex,\mu_2}$ where $h(\mex)$ is independent of the model choice. Therefore, the comparison of two models can be done without computing $h(\mex)$.

\sssu{Comparison of observables average}

Another criterion, easier to compute, is to compare the expected value of the observables average, $\mex[{\cO_k}]$, known from (\ref{AvOk}) to the empirical average 
$\PT{\cO_k}{ \omega}$. Error bars are expected to follow the central limit theorem where fluctuations are given  by eq. (\ref{d2Pbeta}). Examples are given in Fig. 
\ref{monomialaverages}. Note that the comparison of observables average is less discriminant than minimizing the Kullback-Leibler divergence, since there are infinitely many 
possible models matching the observables average.

%% file: Method.tex
\su{Monte-Carlo method for spatio-temporal Gibbs distribution}\label{SMethod}

\ssu{The advantages and limits of transfer matrix method}\label{SLimitsTransfer}

 The advantage of the transfer matrix technique method is that it is mathematically exact: given a
potential $\H$, it gives the Gibbs distribution and topological pressure without computing a partition function; given the parametric form (\ref{H}) where the parameters 
$\beta_k$s has to be determined (``learned''), it provides the unique solution. On numerical grounds, this method provides an optimal estimation, in the limits of the 
error made when observing the observables empirically, this error being characterized by  the central limit theorem. Its main drawback is that the transfer matrix 
$\cL$ has $2^{NR}$ entries ! Although, most of those entries are zero ($2^N$ non zero entries per row, thanks to the compatibility conditions) it is simply too huge 
to handle cases where $NR > 24$.

Focusing thus on the huge number of states in the set of blocks, it is clear that any method requiring the extensive description of the phase space  fails as $NR$
grows. Additionally, while the accessible phase space is huge, the \textit{observed} phase space (e.g. in an experimental raster) is rather small. Several strategies 
exist to avoid the extensive description of the phase space. Here, we propose an approach based on Monte-Carlo sampling. 

The idea is the following. Given a potential $\H$ we find a strategy to approximately compute the average $\moy{\cO_k}$ of observables $\cO_k$ under the Gibbs distribution 
$\mu$, using a statistical Monte-Carlo sampling of the phase space. For that purpose, the algorithm generates a raster following the statistics defined by the potential 
$\H$, and computes the observables on this artificial raster.  Thanks to the estimation of the observables, the parameters of the model ($\beta_k$s) can be found by 
modifying their values to minimize iteratively the distance between the values of the observables estimated on the real raster, and the values estimated with the 
Monte-Carlo sampling. Powerful algorithms exist for that, taking into account the uncertainty on the empirical averages ruled by the central limit theorem 
(\cite{dudik-phillips-etal:04} and \cite{broderick-dudik-etal:07})).

\ssu{The Monte-Carlo-Hastings algorithm}

The Monte-Carlo-Hastings method consists in sampling a target probability distribution $\mu$ by constructing a Markov chain whose invariant probability
is $\mu$ \cite{hastings:70}. The transition probability of this Markov chain, between two states $\om{1}$ and $\om{2}$ is:
\begin{equation}\label{AccProb}
 P[\om{1}|\om{2}] = \max (\frac{Q(\om{1}|\om{2})}{Q(\om{2}|\om{1})}\frac{\moy{\om{2}}}{\moy{\om{1}}},1).
\end{equation}

The function $Q()$ can have different forms, allowing in particular to speed-up the convergence rate of the algorithm. Such specific forms are highly dependent
on the form of $\H$, and there is no general recipe to determine $Q$, given $\H$. The contribution of $Q$ cancels in (\ref{AccProb})  whenever $Q$ is symmetric
($Q(\omega|\omega') = Q(\omega'|\omega$)). We make this assumption in the sequel. Practically, we take $Q$ as the uniform distribution 
corresponding to flipping one spike at each iteration of the method.\\

In classical Monte-Carlo approaches in statistical physics, the normalization factor of the Gibbs distribution, the partition function, cancels when 
computing the ratio of two blocks probabilities $\frac{\moy{\om{2}}}{\moy{\om{1}}}$. The situation is different in the presence of spatio temporal constraints, as
shown in eq. (\ref{P_large_blocks}): ``boundary terms'' $\lpfc{\bloc{0}{D-1}}$, $\rpfc{\bloc{n-D+1}{n}}$ remain. Actually, the same would hold in statistical physics
problem with spatial interactions if one were to compare the probability of bulk spin-chains with distinct boundary conditions. 

This problem can however be circumvented thanks to the following remarks:

\begin{enumerate}

\item If one compares the probability of two blocks $\om{1},\om{2}$ of range $n\geq 2D+1$, with $\blp{1}{0}{D-1}=\blp{2}{0}{D-1}$ and $\blp{1}{n-D+1}{n}=\blp{2}{n-D+1}{n}$ then 
(\ref{P_large_blocks}) reads:
$$
\frac{\moy{\blp{2}{0}{n}}}{\moy{\blp{1}{0}{n}}} = e^{\Delta \H(\om{1},\om{2},0,n)}
$$
with
$$
\Delta \H(\om{1},\om{2},0,n)
=\H\pare{\blp{1}{0}{n}}-\H\pare{\blp{2}{0}{n}}.
$$

Thus, the Monte-Carlo transition probability (\ref{AccProb}) is only expressed as a difference of potential of the two blocks.

\item $\Delta \H(\om{1},\om{2},0,n) = \sum_{k=1}^K \beta_k 
\Delta \cO_k(\om{1},\om{2},0,n)$, with:
$$
\Delta \cO_k(\om{1},\om{2},0,n)
=\sum_{l=0}^{n-D} \bra{
\cO_k\pare{\blp{2}{l}{D+l}}
-
\cO_k\pare{\blp{1}{l}{D+l}}
}.
$$ 
Since the $\cO_k$s are monomials, many terms
 $\cO_k(\omega_l^{'l+D}) - \cO_k(\omega_l^{l+D})$ cancel. Assuming that we flip a spike at position $(k,t)$, $k \in \Set{1,\dots,N}$, $t \in \Set{D,n-D}$, we have indeed:
$$
\Delta \cO_k(\om{1},\om{2},0,n)
=\sum_{l=t-D}^{t} \bra{
\cO_k\pare{\blp{2}{l}{D+l}}
-
\cO_k\pare{\blp{1}{l}{D+l}}
}
$$
Since the difference $\cO_k\pare{\blp{2}{l}{D+l}}
-
\cO_k\pare{\blp{1}{l}{D+l}} \in \{-1,0,1\}$, the computational cost of $\Delta \cO_k(\om{1},\om{2},0,n)$ is minimal if one makes a list of monomials affected by the 
flip of spike $(k,r)$, $r=0, \dots D$.

\end{enumerate}

\ssu{Convergence rate} \label{Sconv}

The goal of Monte-Carlo-Hastings algorithm is to generate a sample of a target probability  obtained by iteration of the Markov chain defined by eq. (\ref{AccProb}). 
In our case, this sample is a raster $\bloc{0}{T-1}$, distributed according to a Gibbs distribution $\mu$. Call $N_{flip}$ the number of iterations (``flips'' in our case)
of the Monte-Carlo algorithm. As $N_{flip} \to +\infty$ the probability that the algorithm generates a raster $\bloc{0}{T-1}$ tends to $\moy{\bloc{0}{T-1}}$. Equivalently,
if one generates $N_{seed}$ rasters and denote $\#\pare{\bloc{0}{T-1}}$ the number of occurrences of a specific bloc $\bloc{0}{T-1}$, then:

$$\lim_{N_{seed} \to +\infty} \lim_{N_{flip} \to +\infty}
\frac{\#\pare{\bloc{0}{T-1}}}{N_{seed}} = \moy{\bloc{0}{T-1}}.$$

The convergence is typically exponential with a rate depending on $\H$.

Now, the goal here is to use a Monte-Carlo raster to
estimate $\moy{\cO_k}$ by performing the empirical average $\PT{\cO_k}{\omega}$  on that raster.
However, as explained in section \ref{SEmpAv}, even if the raster is distributed according to $\mu$ (corresponding thus to taking the limit $N_{flip} \to + \infty$) 
the empirical average $\PT{\cO_k}{\omega}$ is not equal to $\moy{\cO_k}$, it converges to $\moy{\cO_k}$ as $T \to +\infty$, with an exponential rate (see footnote 
\ref{frate}). More precisely, the probability that the difference $\abs{\PT{\cO_k}{\omega} - \moy{\cO_k}}$ exceeds some $\epsilon >0$ behaves like
$ \exp(-T \times I(\epsilon))$ where $I(\epsilon)$, called large-deviations rate, is the Legendre transform of the topological pressure \cite{chazottes-keller:09}.

When $T$ is large we have:
\beq\label{GaussDev}
\moy{\abs{\PT{\cO_k}{\omega} - \moy{\cO_k}}> \epsilon}
 \simeq \exp(\frac{-T \times {\epsilon}^2}{\sigma(\cO_k)})
\eeq
where $\sigma(\cO_k)=\sqrt{\moy{\cO_k}(1-\moy{\cO_k})}$
is the mean-square deviations of $\cO_k$.

As a consequence, to obtain the exact average $\moy{\cO_k}$
from our Monte-Carlo algorithm we would need to take the limits:
\beq\label{limits}
\lim_{T \to +\infty} \lim_{N_{seed} \to +\infty} \lim_{N_{flip} \to +\infty}
\frac{\#\pare{\bloc{0}{T-1}}}{N_{seed}},
\eeq
in \textit{that order}: they do not commute. A prominent illustration of this point is illustrated in fig. \ref{error_vs_Ntimes}.\\

For notation homogeneity we note from now on $T-1 \equiv N_{times}$ for the raster length. When dealing with numerical simulations with a finite number of sample, 
the goal is to minimize the probability that the error is bigger than a real number $\epsilon$,
by suitable choice of:

\begin{itemize}
 \item The raster length: $T-1 = N_{times}$.
 \item The number of flips: $N_{flip}$.
 \item The number of seed: $N_{seed}$.
\end{itemize}

Let us now establish a few relations between those parameters. First, it is somewhat evident that $N_{flip}$ must be at least proportional to $N \times N_{times}$ in order to give a chance to all spikes in the raster to be flipped at least once. This criterion respects the order of limits in (\ref{limits}). 

Since $\mu$ is ergodic one can in principle estimate the average of observables by taking $N_{seed}=1$ and taking $N_{times}$ large. However, the larger $N_{times}$ the larger $N_{flip}$ and too big $N_{times}$ leads to too long simulations.  On the opposite, one could generate a large number $N_{seed}$ of raster with a small $N_{times}$. This would have the advantage of reducing $N_{flip}$ as well.
However, the error (\ref{GaussDev}) would then be too large. So, one needs to find a compromise: $N_{times}$ large enough to have small Gaussian fluctuations (\ref{GaussDev}) and small
enough to limit $N_{flip}$. Then, by increasing $N_{seed}$,
one approaches the optimal bound on fluctuations given by (\ref{GaussDev}). Additionally, this provides error bars. 

%% file: Tests.tex
\section{Numerical tests }\label{STests}


In this section, performance in terms of convergence rate and CPU time for increasing values of $N$ (number of neurons) are discussed. First, we consider potentials (\ref{H})  where 
the $\cO_k$s are observables of the form (\ref{DefObs}), ("polynomial potentials") and we compare the Monte-Carlo results to those obtained using the transfer matrix and Perron - Frobenius theorem. 
As discussed in section \ref{SLimitsTransfer} the transfer matrix method becomes rapidly numerically intractable, so that the comparison between Monte-Carlo averages 
and exact averages cannot be done for large $N$. To circumvent this problem, we introduce, in section \ref{SExample} a specific class of potentials for which the 
analytical computation of the topological pressure as well as observable averages can be analytically done, whatever $N,R$. This provides another series of tests.      

\ssu{Polynomial potentials}

In this section, we present Monte-Carlo simulations with a potential of the form (\ref{H}) where the $\cO_k$s are
100 observables randomly chosen among the $2^{NR}$ possibilities. More precisely, 
we select randomly a fraction $\frac{1}{R}$ of "rates",
a fraction $\frac{1}{R}$ of pairwise terms and so on.

\sssu{Checking observables average}

\begin{figure}[ht!]
\begin{center} 
\includegraphics[height=6cm,width=10cm]{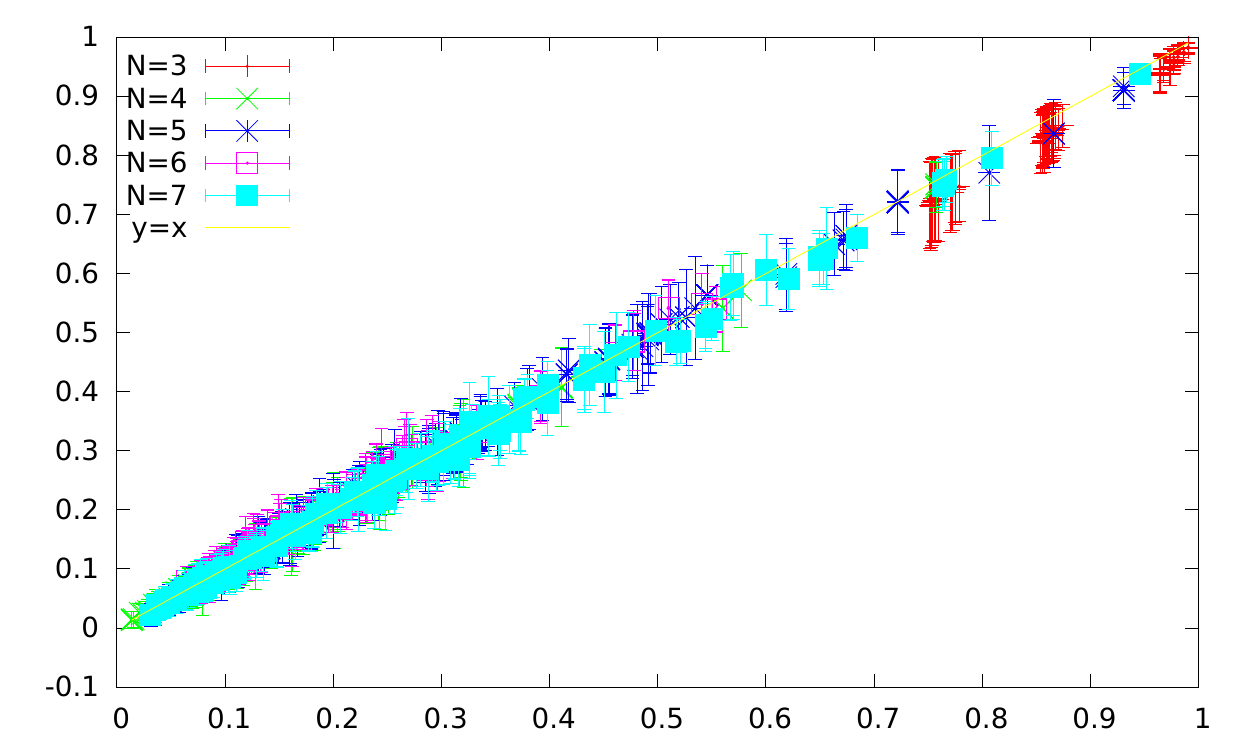} 
\end{center}
\caption{Comparison between the estimated and real values of observable averages.}
\label{monomialaverages}
\end{figure}

The figure \ref{monomialaverages} shows the comparison between the exact values of the observable averages (ordinate) and the estimated Monte-Carlo values (abscissa). 
The error bars have been computed with $N_{seed}$ samples. The tests were performed with $N_{seed}=20$, $N_{times} = 10000$ and $N_{flip} = 100000$. 
In this example, $N$ goes from $3$ to $8$ and $R=3$. For larger values of $NR$ the numerical computation with the transfer matrix method is not possible any more 
($NR=24$ corresponds to matrices of size $16777216 \times 16777216$).

\sssu{Convergence rate}

\begin{figure}[h!]
\begin{center} 
\includegraphics[height=6cm,width=6cm]{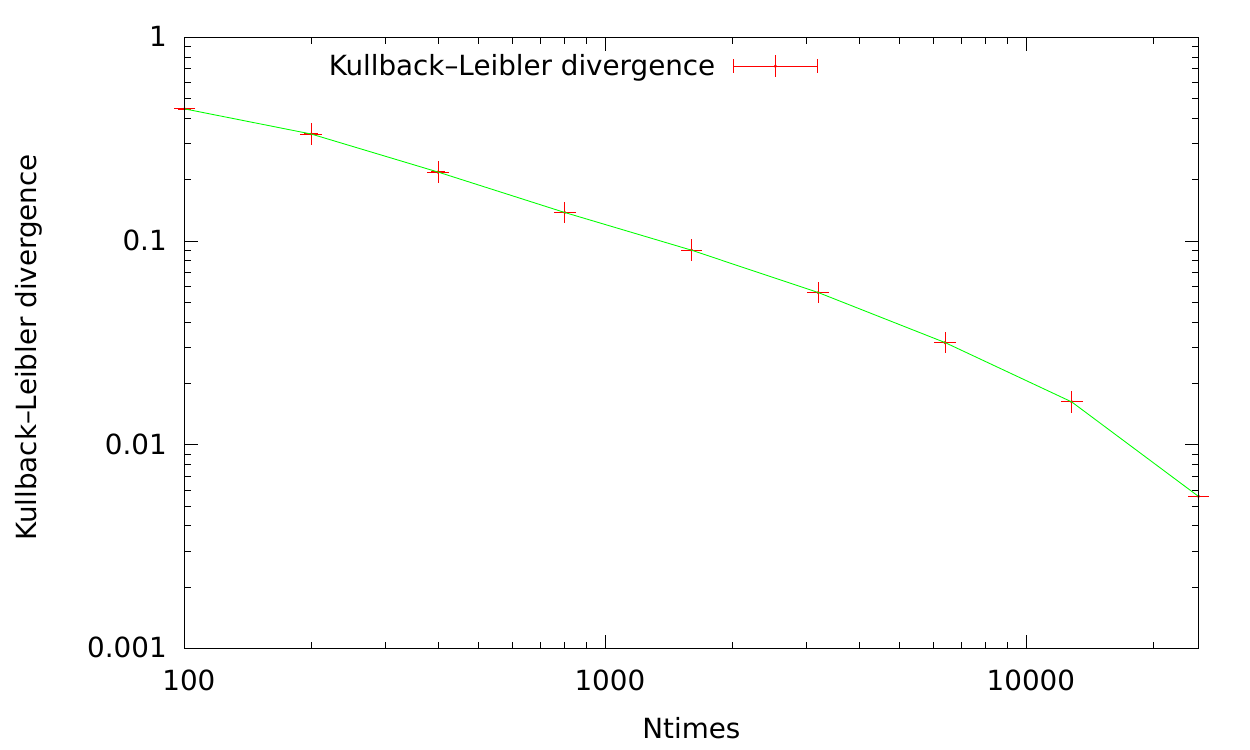} 
\end{center}
\caption{Evolution of Kulback-Leibler divergence \ref{dKL} as a function $N_{times}$.}
\label{dkl_ntimes}
\end{figure}

In this section, we show how the Kulback-Leibler divergence varies as a function of $N_{times}$. 
The figure \ref{dkl_ntimes} shows the evolution of the Kulback-Leibler divergence between the real distribution and its estimation with
the Monte-Carlo method.

We also consider the error
\begin{equation}\label{relative_error}
 error = \max_{k=1 \dots K} \abs{1-\frac{\PT{\cO_k}{ }}{\moy{\cO_k}}}.
\end{equation}
 As developed in section \ref{Sconv}, this quantity is expected to converge to $0$ if 
$N_{times} \to +\infty$ when $N_{flip}$ grows proportional to $N \times N_{times}$. For finite $N_{times}$, $N_{flip} \to +\infty$ the error is controlled by 
the central limit theorem. The probability that the error on the average of observable $\cO_k$ is bigger than $\epsilon$ (eq. (\ref{GaussDev})), 
behaves like $\exp\pare{\frac{-N_{times} \times {\epsilon}^2}{\sigma(\cO_k)}}$.

\begin{figure}[h!]
\begin{center} 
\includegraphics[height=6cm,width=5cm]{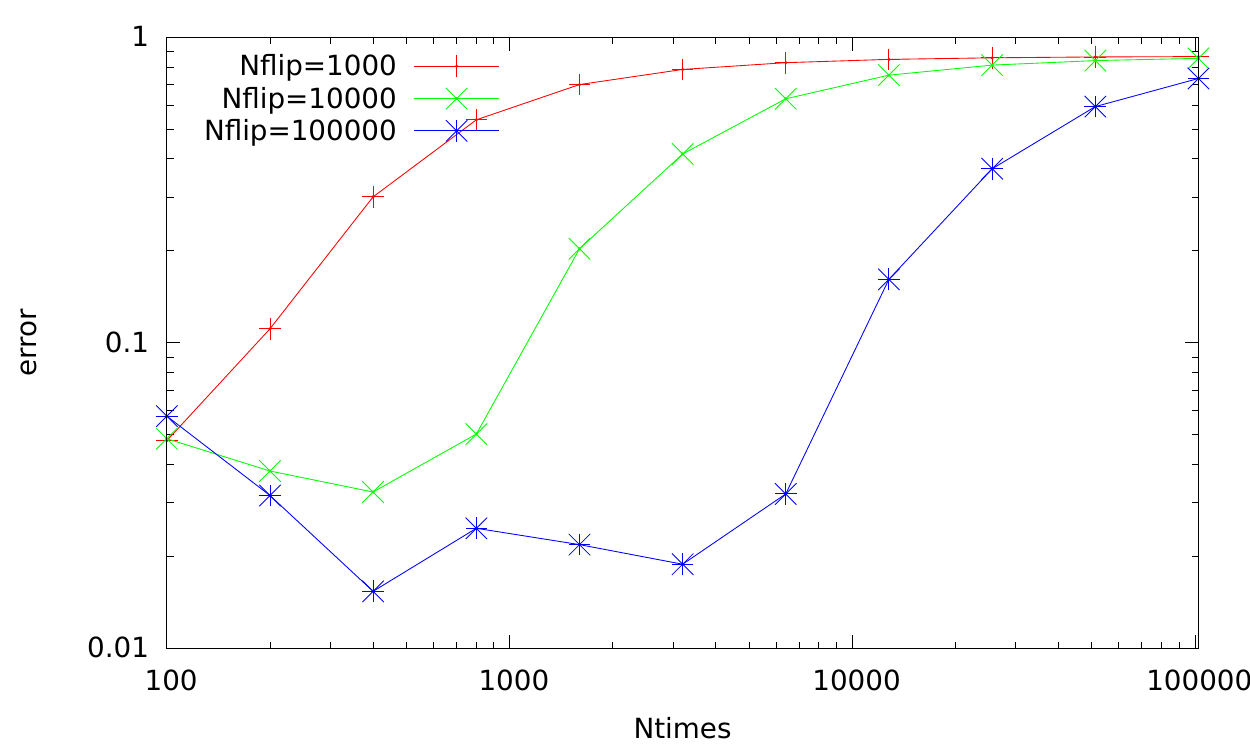} 
\includegraphics[height=6cm,width=5cm]{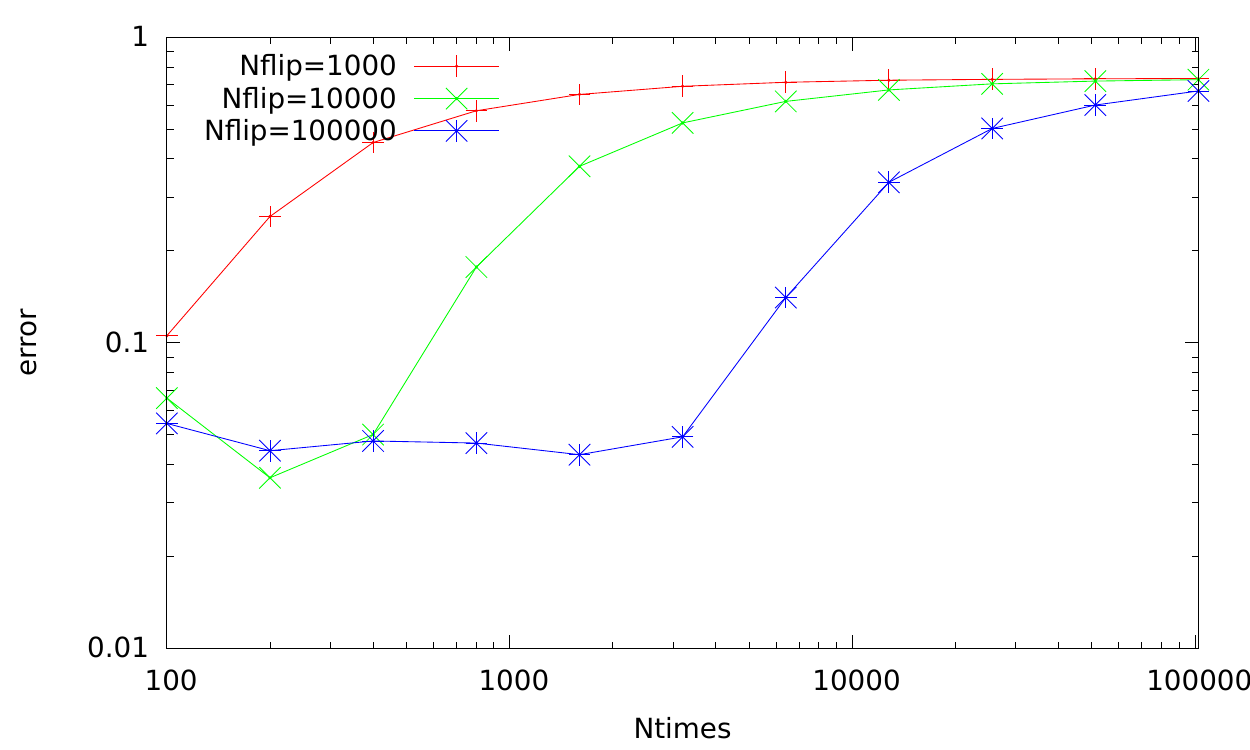} 
\end{center}
\caption{Error as a function of $N_{times}$, for several values of $N_{flip}$ ($1000$, $10000$, $100000$). (left) $N=3$; (right) $N=7$.}
\label{error_vs_Ntimes}
\end{figure}

 On the opposite, if $N_{flip}$ stays constant while $N_{times}$ grows, the error is expected to first decrease up to a minimum after which it increases. This is 
because the number of flips is insufficient to reach the equilibrium distribution  of the Monte-Carlo-Hastings Markov chain. This effect is presented in fig. 
\ref{error_vs_Ntimes}. It  shows the error (\ref{relative_error}) as a function of $N_{times}$ for $N=3$ to $N=7$ neurons with $3$ different $N_{flip}$ values ($1000$, $10000$, $100000$). 
 
Clearly, the number of flips $N_{flip}$ should be at least more than $N \times N_{times}$ in order to give a chance to all spikes in the raster to be flipped at least once. A value $N_{flip}=10 \times N \times N_{times}$ seems to be enough and computationally reasonable to perform the estimations. 
With an $N_{seed}=20$, we have results with a reasonable error around the mean values.

\pagebreak
   
\sssu{CPU time}

\begin{figure}[h!]
\begin{center} 
\includegraphics[height=6cm,width=5cm]{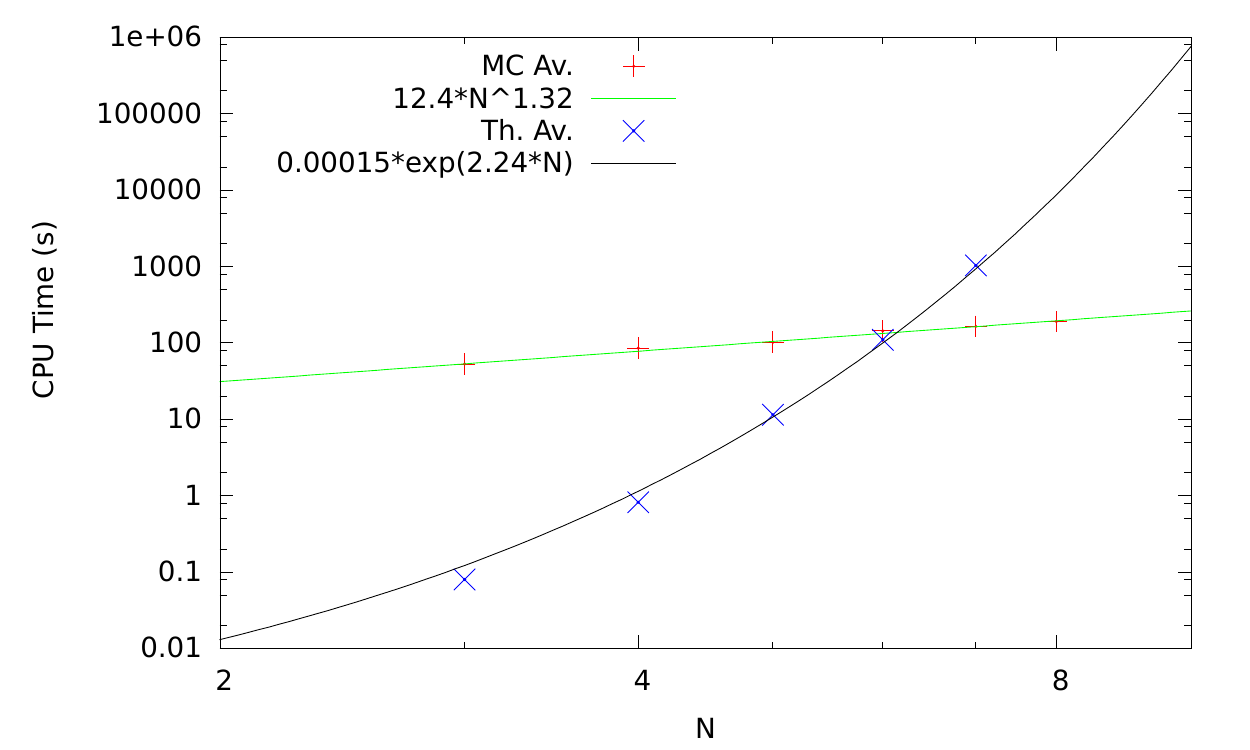} 
\end{center}
\caption{The CPU time necessary to obtain the observable averages presented in fig. \ref{monomialaverages}, for the Monte-Carlo Average (MC Av.) and for the exact average
 (Th. Av.), as a function of $N$. The full lines correspond to fit.}
\label{CPU_vs_Ntimes}
\end{figure}

Here we compare the CPU time for a Monte-Carlo simulation and the time for a computation with the transfer matrix. The figure \ref{CPU_vs_Ntimes} illustrates this.
We have plotted the CPU time necessary to obtain the observables average presented in fig. \ref{monomialaverages}, for the Monte-Carlo Average and for the exact average,
as a function of $N$. The CPU time for Monte-Carlo increases slighty more than linearly while the CPU time for transfer matrix method increases exponentially fast: 
note that $R=3$ here so that $R\log 2=2.08$, close from the exponential rate found by fit: $2.24$.

\begin{figure}[h!]
\begin{center} 
\includegraphics[height=6cm,width=5cm]{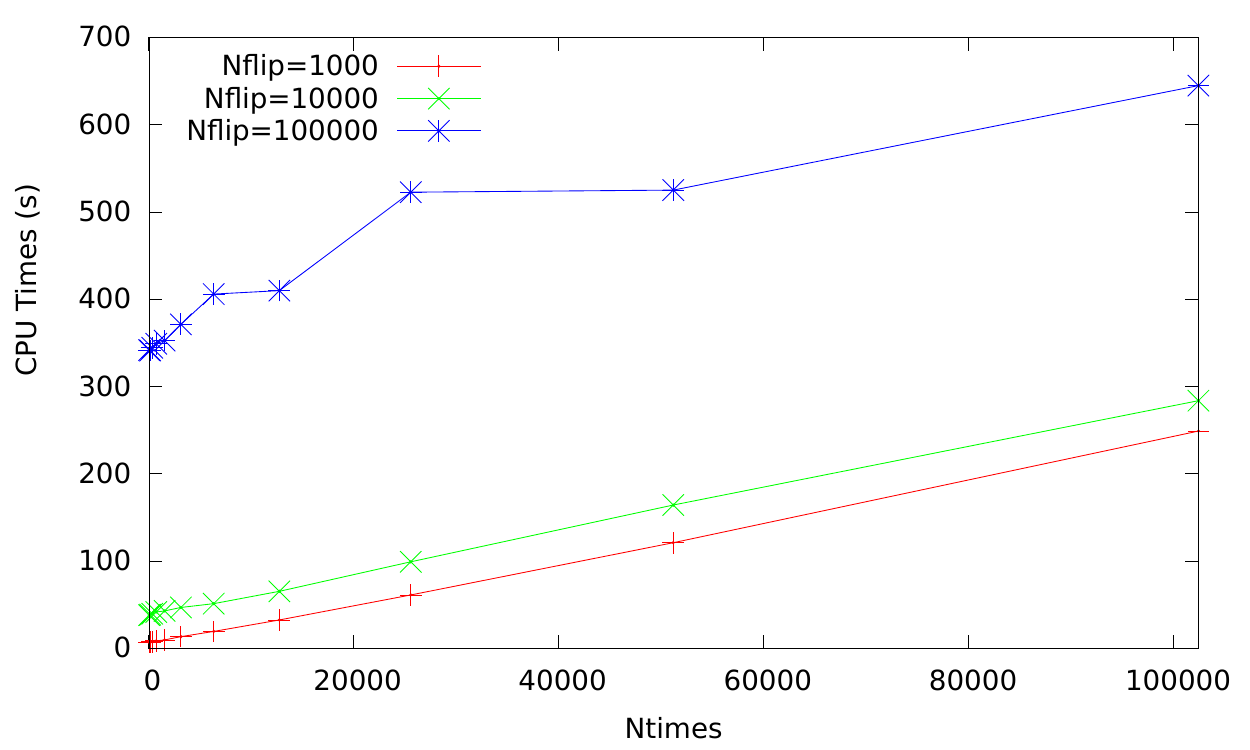} 
\includegraphics[height=6cm,width=5cm]{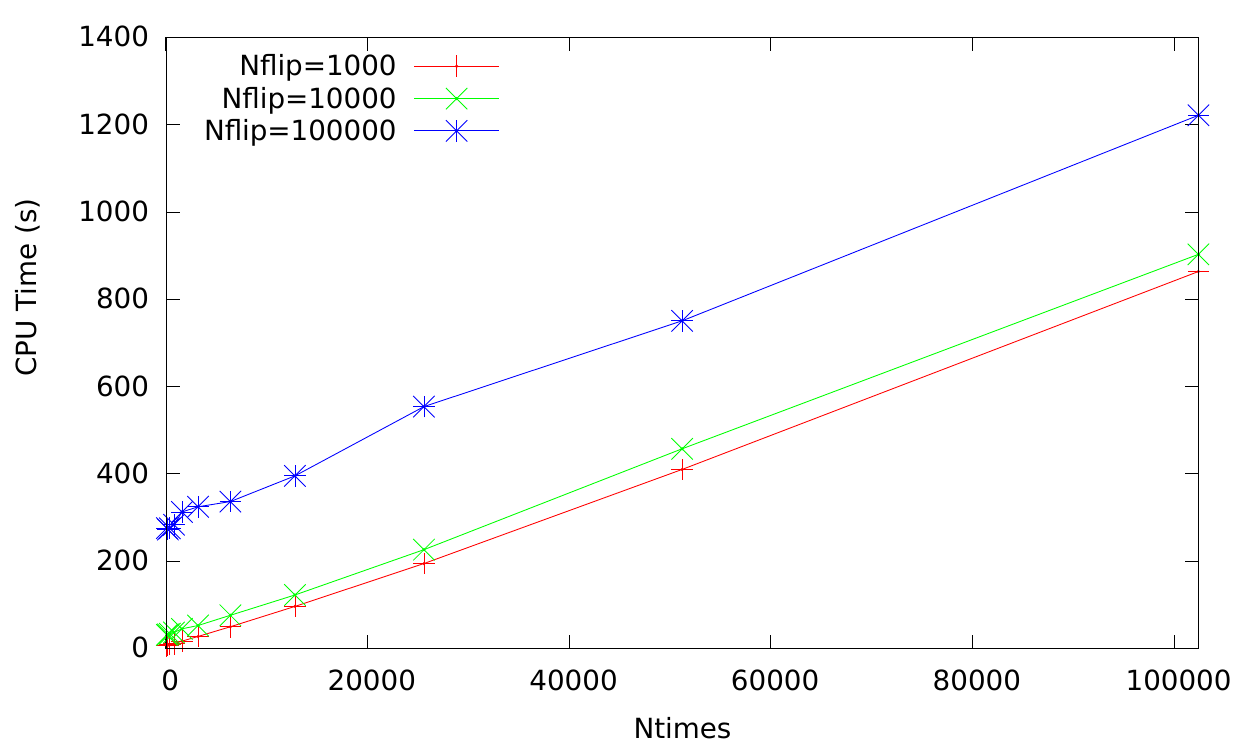} 
\end{center}
\caption{The CPU time ($T_{cpu}$) as a function of $N_{times}$. $T_{cpu}$ increases in a lnear fashion with $N_{time}$ as 
$a \times N + b$ where $a$ is a function of $N$ and $b$ is a function of $N_{flip}$.}
\label{time_vs_Ntimes}
\end{figure}

We also plot in fig. \ref{time_vs_Ntimes} the CPU times as a function of $N_{times}$ with three $N_{flip}$ values (1000, 10000, 100000), for $3$ and $7$ neurons. The CPU time increases in a linear
fashion with the $N_{times}$ value. The CPU time also increases linearly with the $N_{flip}$ value (for the same $N$ value). The simulations have been done on a computer 
with the following  characteristics: 7 Intel(R) Xeon(R) 3.20GHz processors with a 31.5 Gb RAM.

\begin{figure}[h!]
\begin{center} 
\includegraphics[height=6cm,width=5cm]{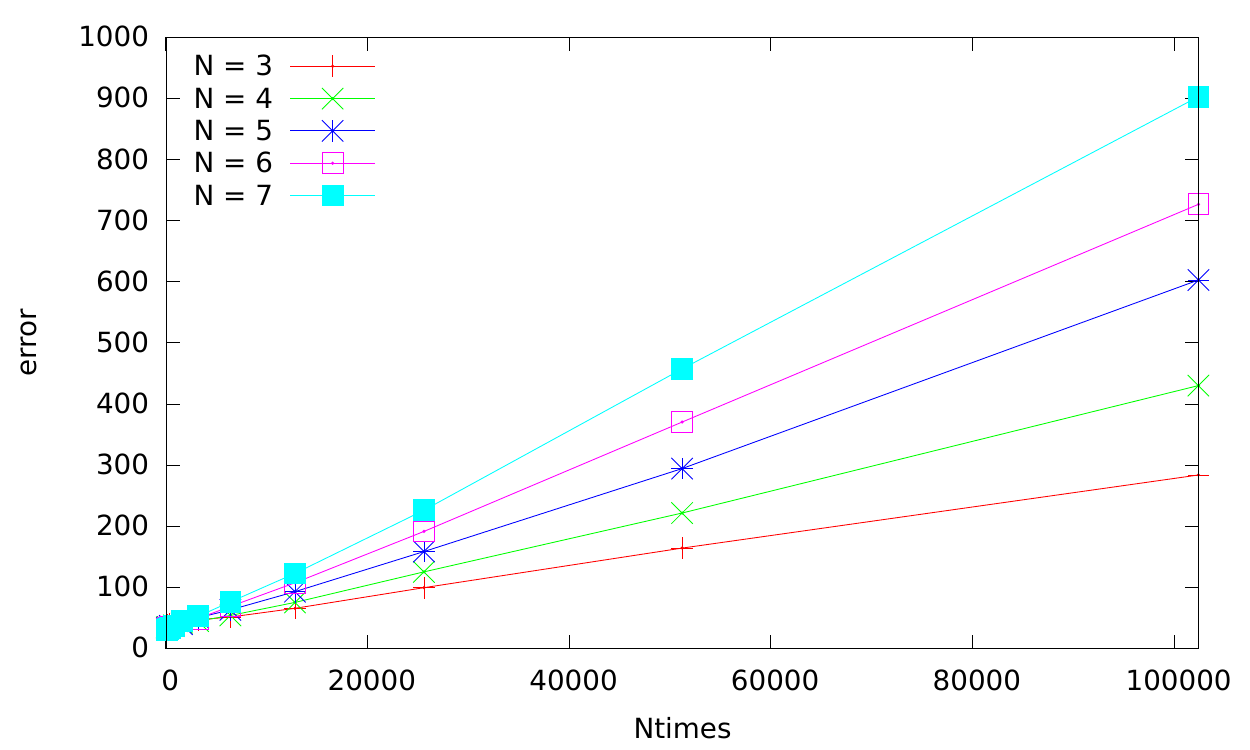} 
\end{center}
\caption{The CPU time ($T_{cpu}$) as a function of $N_{times}$ for several $N$ values ($Nflip = 10000$).}
\label{time_vs_Ntimes_AllN}
\end{figure}

The figure \ref{time_vs_Ntimes_AllN} compares the CPU time increase with $N_{times}$ for several $N$ values. It show that the CPU time increases
linearly with $N_{times}$ (as figure \ref{time_vs_Ntimes}). However, the slope increases with the number of neurons $N$.

\pagebreak

\input{Example}

%% file: Example.tex
\ssu{An analytically solvable example as a benchmark}\label{SExample}

In this section we consider a specific example of potentials  for which the topological pressure is analytically computable, whatever $N,R$. As a consequence the 
average of observables and fluctuations can also be computed. This example is obviously rather specific, but its main interests are to provide a didactic illustration
 of thermodynamic formalism application as well as a benchmark for numerical methods.

\sssu{Analytical setting}

We fix the number of neurons $N$ and the range $R$ and we choose $L$ distinct pairs $(i_l,t_l)$, $l=1 \dots L$,
$i_l \in \Set{1, \dots, N}$, $t_l \in \Set{0, \dots, D-1}$. To this set is associated a set of $K=2^L$ events
$\cE_k=\pare{\omega_{i_1}(t_1), \dots, \omega_{i_l}(t_l)}, \, k=0, \dots, K-1$.
For example, if $L=2$, there are $4$ possible events
$\cE_0=(0,0)$ : neuron $i_1$ is not firing at time $t_1$ and neuron $i_2$ is not firing at time $t_2$;
$\cE_1=(0,1)$ : neuron $i_1$ is not firing at time $t_1$ and neuron $i_2$ is firing at time $t_2$;
and so on. It is convenient to have a label $k$ corresponding to the binary code of the event.

We define $K$ observables $\cO_k$ of range $D$ taking binary values $0,1$. For a block $\bloc{0}{D}$, $\cO_k\bra{\bloc{0}{D}}=0$ if the event $\cE_k$ is not
realized in the bloc   $\bloc{0}{D}$ and is $1$ otherwise. In the example above, $\cO_0\bra{\bloc{0}{D}}=1$ if neuron $i_1$ is not firing at time $t_1$ and neuron 
$i_2$ is not firing at time $t_2$
in the block $\bloc{0}{D}$. Thus, for $N=3$, $R=4$, $(i_1,t_1)=(1,0); (i_2;t_2)=(1,1)$, 

$$
\cO_0\bra{\tiny{\pare{
\begin{array}{cccccc}
0 & 0 & 0 & 1\\
0 & 1 & 0 & 1\\
0 & 1 & 0 & 1\\
\end{array}
}}}=
\cO_0\bra{\tiny{\pare{
\begin{array}{cccccc}
0 & 0 & 0 & 1\\
1 & 1 & 0 & 1\\
1 & 1 & 0 & 1\\
\end{array}
}}}=1, 
$$
while
$$
\cO_0\bra{\tiny{\pare{
\begin{array}{cccccc}
1 & 0 & 0 & 1\\
0 & 1 & 0 & 1\\
0 & 1 & 0 & 1\\
\end{array}
}}}=
\cO_0\bra{\tiny{\pare{
\begin{array}{cccccc}
0 & 1 & 0 & 1\\
1 & 1 & 0 & 1\\
1 & 1 & 0 & 1\\
\end{array}
}}}=0.
$$

We finally define a potential $\cH$ as in (\ref{H}), $\H \, = \, \sum_{k=1}^K \beta_k \cO_k$.\\

For this type of potentials, whatever $\bloc{0}{D-1}$,
$\sum_{\bloc{0}{D-1}} e^{\H\pare{\bloc{0}{D}}}$
is a independent of $\omega(D)$. As a consequence
of the Perron-Frobenius theorem $\sb=\sum_{\bloc{0}{D-1}} e^{\H\pare{\bloc{0}{D}}}$ and therefore:
$$
\cP(\H) = (N-L)\log(2) + \log\bra{\sum_{k=1}^K e^{\beta_k}}. 
$$

As a consequence, from (\ref{AvOk}), giving the average of observable $\cO_k$ as the derivative of $\cP$ with respect to $\beta_k$:
$$
\moy{\cO_k}=\frac{e^{\beta_k}}{\sum_{n=1}^K e^{\beta_n}}
$$

The fluctuations of observables can also be estimated as well. They are Gaussian with a covariance matrix given by the Hessian of $\cP$ (see eq. (\ref{d2Pbeta})) 
and the central limit theorem. \\

\textbf{Remarks.} 

\bit

\item An important assumption here is that observables do not depend on $\omega(D)$. This important simplification as well as the specific form of observables makes 
the computation of $\cP$ tractable. Note however that, although $\cH$ does not depend on $\omega(D)$ as well, the \textit{normalized} potential
and therefore the conditional probability $\Probc{\omega(D)}{\bloc{0}{D-1}}$  depend on $\omega(D)$  thanks to the normalization factor $\cG$ and
its dependence in the right eigenvector $R$ of the Perron-Frobenius matrix.


\eit

\sssu{Numerical results for large scale networks}

Let us now use this example as a benchmark for our Monte-Carlo method. We have considered a case with range $R=4$ and $L=6$ pairs, corresponding to $2^6=64$ terms in 
the potential. We have analyzed the convergence rate as a function of $N$ the number of neurons and $N_{times}$, the time length of the Monte-Carlo raster. The number
of flips, $N_{flip}$ is fixed to $10\times N \times N_{times}$, so that, on average, each spike of the Monte-Carlo raster is flipped 10 times in one trial.
  
\begin{figure}[h!]
\begin{center} 
\includegraphics[height=7cm,width=6.5cm]{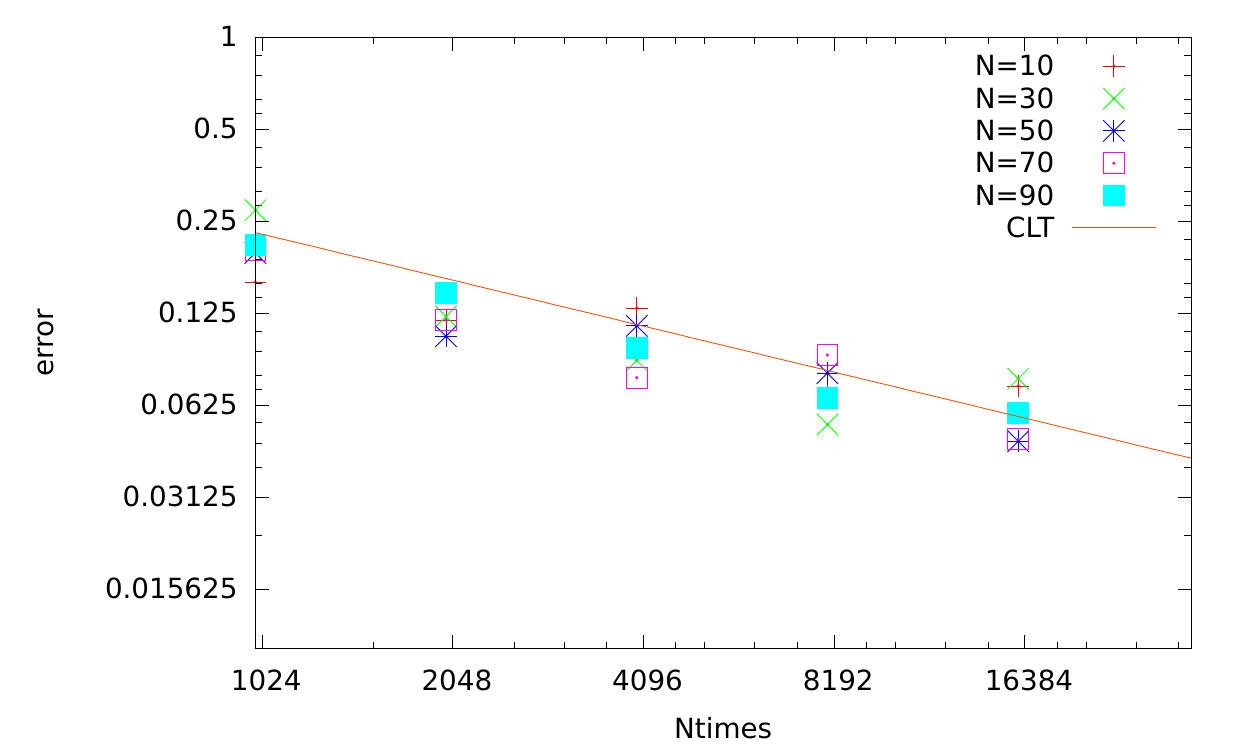} 
\includegraphics[height=7cm,width=6.5cm]{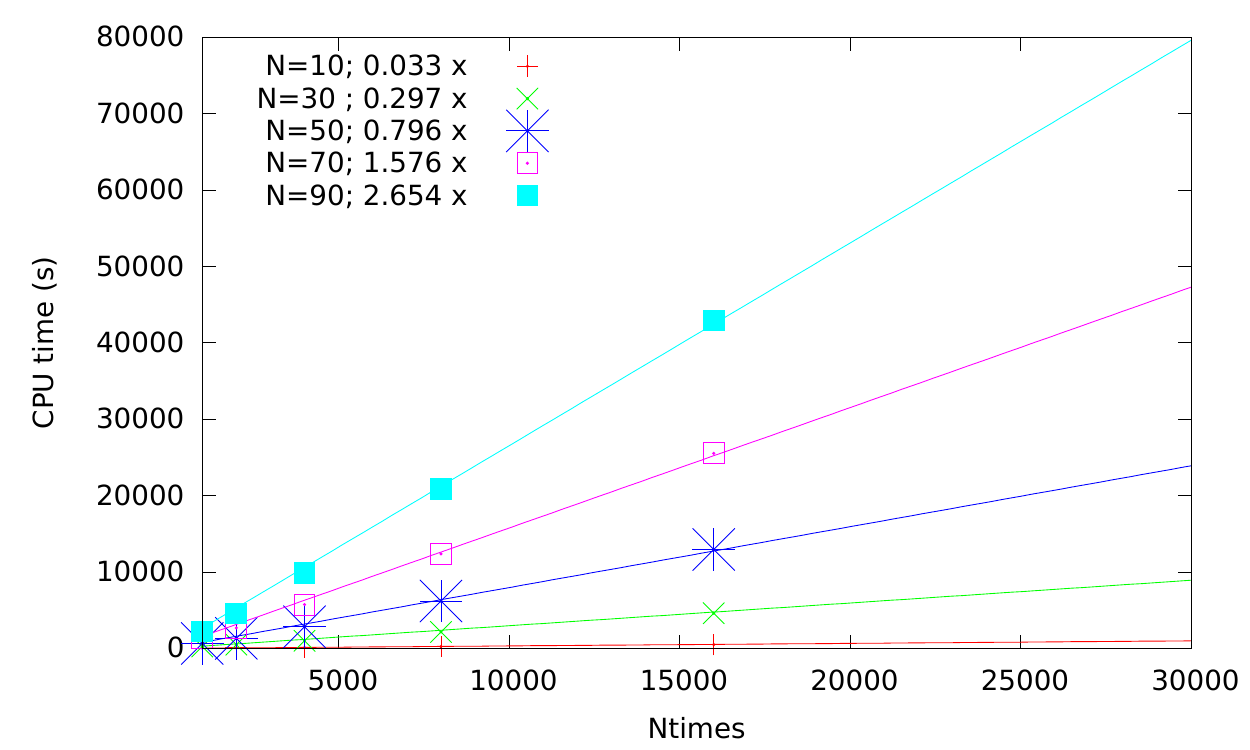} 
\end{center}
\caption{(Left). Relative error as a function of $N_{times}$ for several values of $N$. (CLT) indicates the decay expected from the central limit theorem. As expected 
from the Central Limit Theorem , the error decreases as power law $N_{times}^{-\frac{1}{2}}$. (Right). CPU time as a function of $N_{times}$ for several values of $N$. It increases linearly as $CPU=a \times N_{times}$ In the legend, next to the value of $N$ 
we indicate the value of $a$.}
\label{FError_MC_analytic_L6}
\end{figure}

In fig. \ref{FError_MC_analytic_L6} (left) we have shown the relative error (eq. \ref{relative_error})
as a function of $N_{times}$ for several values of $N$.
The empirical average $\PT{\cO_k}{ }$ is computed on $10$ Monte-Carlo rasters. That's why we don't write the index $\omega$ in the empirical probability. 
We stopped the simulation when the error is lower than $5 \%$. As expected from the Central Limit Theorem (CLT), the error decreases as a power law 
$C N_{times}^{-\frac{1}{2}}$ where the constant $C$ has been obtained by fit.

In fig. \ref{FError_MC_analytic_L6} (right) we have drawn the CPU time as a function of $N_{times}$ for several values of $N$. It increases linearly with $N_{times}$,
with a coefficient depending on $N$. The simulation is relatively fast: it takes $2h30$ for $N=60$, $N_{times}=8000$ (with $10$ Monte-Carlo trials) on a
7 processors machine (each of these processors has the following specifications: Intel(R) Xeon(R) CPU 2.27GHz, 1.55 MB of cach memory size) with a 17.72 GB RAM.

%% file: Conclusion.tex
\section{Discussion and perspectives}

In this paper, we have shown how maximum entropy models can be extended to analyse the spatio-temporal dynamics of the neural activity. This raises specific issues, 
mainly related to the fact that the normalization of the Gibbs potential depends on the past activity. We have shown that transfer matrices results allow to handle 
this problem properly, providing additionally crucial information on statistics (especially average of observables and fluctuations of empirical averages). The 
challenge is then to be able to fit these models to the recordings. A major step in the fitting process is to compute the observables generated by the model for a given 
set of parameters. We have proposed a first method, based on the transfer matrix. It gives exact results, but can only be applied to small subsets of neurons. We have 
then designed a Monte-Carlo approach that overcomes this issue, confirmed by several tests. 
 
In fact, matching Gibbs averages of observables is only a first, although crucial, step towards spike train analysis of neuronal activity. The next step consists of 
fitting the parameters of a model from an experimental raster. Basically, this corresponds to minimizing the Kullback-Leibler divergence (\ref{dKL}) between the model
and the empirical measure. We have reviewed some possible techniques in the section \ref{SInfer}. The application of our 
method to fit Gibbs distributions on large-scale retina recordings will be considered in a forthcoming paper.\\

As a final issue we would now like to discuss shortcomings of Maximum entropy Models.
Although initially proposed as powerful methods for neuroscience applications, many future reports have cast
doubt on how useful (spatial pairwise) Maximum Entropy models were. These criticism include the role of common input \cite{macke-cunningham-etal:11}, the role
of higher order correlations \cite{staude-gruen-etal:10,ohiorhenuan-mechler-etal:10}, scaling properties \cite{roudi-nirenberg-etal:09} and non stationarity \cite{staude-gruen-etal:10}. As a matter of fact, the models presented in the section \ref{SOther} are clear and efficient alternatives to spatial pairwise maximum entropy models. Let us now comment these shortcomings.\\

First, as we have developed, the maximum entropy approach is definitely not limited to spatial pairwise interactions. Especially, the role of higher order interactions beyond pairwise equal time ones, provides a clear motivation for including longer temporal history in statistical models of neural data. 

This raises however the question of how one chooses the potential. As exposed in section \ref{SMot} the possible number of constraints is simply overwhelming and one has to make 
choices to reduce their number. These choices can be based on ad hoc assumptions (e.g. rates or instantaneous pairwise correlations are essential in neuronal 
coding) or on empirical constraints (type of cells, spatial localisation). However, this combinatorial complexity is clearly a source of troubles and questions about 
the real efficiency of the maximum entropy problem. This problem is particularly salient when dealing with temporal interactions of increasing memory: even the
number of possible pairwise interactions might be too large to fit all of them on finite size recordings. Additionally, using a too complex potential increases the number of parameters 
necessary to fit the data beyond what the number of available samples allows.

One solution is to try to infer the form of the potential from the data set. Important theorems in ergodic theory \cite{pollicott-weiss:03} as well
as in Variable Length Markov Chains Estimations can be used \cite{buehlmann-wyner:99}. This will be developed in a separate paper. 

It is however quite restrictive to stick to potentials expressed as linear combinations of observables, like (\ref{H}). This form has its roots in thermodynamics and
 statistical physics, but is far from being the most general form. Nonlinear potentials such as (\ref{RateGLM}) (GLM) can be considered as well. Although such potentials can be expressed in the 
form (\ref{H}) from the Hammersley-Clifford theorem \cite{hammersley-clifford:71}, this representation induces a huge redundancy in the coefficients $\beta_k$. 
Examples are known of non linear potentials with relatively small number of parameters $\lambda_l$, which, expressed in the form (\ref{H}), give rise to $2^{NR}$
 parameters $\beta_k$s, all of them being functions of the $\lambda_l$s: see
\cite{cessac:11a,cessac:11b}. Such potentials constitute relevant alternatives to (\ref{H}) where
the formalism described here fully applies.

More generally, alternatives to maximum entropy models  consider different models trying to mimic the origin of the observed correlations. This is the case of the model proposed by \cite{macke-etal:11} when common inputs are added to account for the instantaneous correlations and the GLM model where the numbers of parameters is only $N^2$. However, note that in all
these cases, the models constrain the correlations to be in a specific form, and might not be a good description of the activity either. Testing these models on data is the 
only way to distinguish the most relevant ones. Note that the discrepancy between model and data will probably be more and more obvious with a larger set of neurons.
The validity of a model will also depend on size of the recorded population.\\

Another, even deeper, question is the translation-invariance assumption intrinsic to the maximum entropy principle. When dealing e.g. to transient responses to temporary
stimuli this assumption is clearly highly controversial. Note however that although the maximum entropy principle does not extend to non translation-invariant 
statistics, the concept of Gibbs distribution extend to that case \cite{georgii:88}. Here, Gibbs distributions are constructed via transition probabilities, possibly 
with an infinite memory. Examples of applications to neuronal networks can be found in  \cite{cessac:11a,cessac:11b}. However, the application of this concept to 
analyzing real data, especially the problem of parameters estimation, remains to our knowledge an open challenge.



%% file: MonteCarlo.bbl
\begin{thebibliography}{10}

\bibitem{abeles:82}
M.~Abeles.
\newblock {\em Local Cortical Circuits: An Electrophysiological study}.
\newblock Springer, Berlin, 1982.

\bibitem{ackley-etal:85}
H.~Ackley, E.~Hinton, and J.~Sejnowski.
\newblock A learning algorithm for boltzmann machines.
\newblock {\em Cognitive Science}, pages 147--169, 1985.

\bibitem{peterson-anderson:87}
H.~Ackley, E.~Hinton, and J.~Sejnowski.
\newblock a mean field theory learning algorithm for neural network.
\newblock {\em Complex systems}, pages 995--1019, 1987.

\bibitem{pillow-ahmadian-etal:11}
Yashar Ahmadian, Jonathan~W. Pillow, and Liam Paninski.
\newblock {Efficient Markov Chain Monte Carlo Methods for Decoding Neural Spike
  Trains}.
\newblock {\em Neural Computation}, 23(1):46--96, January 2011.

\bibitem{beck-schloegl:95}
C.~Beck and F.~Schloegl.
\newblock {\em Thermodynamics of Chaotic Systems: An Introduction}.
\newblock Cambridge University Press, Cambridge, 1995.

\bibitem{brillinger:88}
D.~R. Brillinger.
\newblock Maximum likelihood analysis of spike trains of interacting nerve
  cells.
\newblock {\em Biol Cybern}, 59(3):189--200, 1988.

\bibitem{brillinger:92}
D.~R. Brillinger.
\newblock {Nerve Cell Spike Train Data Analysis - a Progression of Technique}.
\newblock {\em J Amer Statist Assn}, 87(418):260--271, 1992.

\bibitem{broderick-dudik-etal:07}
Tamara Broderick, Miroslav Dudik, Gasper Tkacik, Robert~E. Schapire, and
  William Bialek.
\newblock Faster solutions of the inverse pairwise ising problem.
\newblock {\em Submitted (see http://arxiv.org/abs/0712.2437)}, 2007.

\bibitem{brown-barbieri-etal:03}
E.~N. Brown, R.~Barbieri, U.~T. Eden, and L.~M. Frank.
\newblock {Likelihood methods for neural spike train data analysis}.
\newblock {\em Computational Neuroscience: A Comprehensive Approach}, 2003.

\bibitem{buehlmann-wyner:99}
Peter Buehlmann and Abraham~J. Wyner.
\newblock Variable length markov chains.
\newblock {\em The Annals of Statistics}, 27(2), 1999.

\bibitem{cessac:11a}
Bruno Cessac.
\newblock A discrete time neural network model with spiking neurons ii.
  dynamics with noise.
\newblock {\em J. Math. Biol.}, 62:863--900, 2011.

\bibitem{cessac:11b}
Bruno Cessac.
\newblock Statistics of spike trains in conductance-based neural networks:
  Rigorous results.
\newblock {\em Journal of Mathematical Neuroscience}, 1(8), 2011.

\bibitem{chazottes-keller:09}
J.R. Chazottes and G.~Keller.
\newblock Pressure and equilibrium states in ergodic theory.
\newblock {\em Israel Journal of Mathematics}, 131(1), 2008.

\bibitem{chichilnisky:01}
E.~J. Chichilnisky.
\newblock A simple white noise analysis of neuronal light responses.
\newblock {\em Network: Comput. Neural Syst.}, 12:199--213, 2001.

\bibitem{cocco-leibler-etal:09}
Simona Cocco, Stanislas Leibler, and R{\'e}mi Monasson.
\newblock Neuronal couplings between retinal ganglion cells inferred by
  efficient inverse statistical physics methods.
\newblock {\em PNAS}, 106(33):14058--14062, 2009.

\bibitem{cofre-cessac:12}
Rodrigo Cofr\'e and Bruno Cessac.
\newblock Dynamics and spike trains statistics in conductance-based
  integrate-and-fire neural networks with chemical and electric synapses.
\newblock {\em Chaos, Solitons and Fractals, submitted}, 2012.
\newblock submitted.

\bibitem{cornfeld-fomin-sinai:82}
I.~P. Cornfeld, S.~V. Fomin, and Ya.~G. Sinai.
\newblock {\em Ergodic Theory}.
\newblock Springer, Berlin, Heidelberg, New York, 1982.

\bibitem{dudik-phillips-etal:04}
Miroslav Dud\'ik, Steven~J. Phillips, and Robert~E. Schapire.
\newblock Performance guarantees for regularized maximum entropy density
  estimation.
\newblock In {\em Proceedings of the 17th Annual Conference on Computational
  Learning Theory}, 2004.

\bibitem{ganmor-segev-etal:11a}
Elad Ganmor, Ronen Segev, and Elad Schneidman.
\newblock The architecture of functional interaction networks in the retina.
\newblock {\em The journal of neuroscience}, 31(8):3044--3054, 2011.

\bibitem{ganmor-segev-etal:11b}
Elad Ganmor, Ronen Segev, and Elad Schneidman.
\newblock Sparse low-order interaction network underlies a highly correlated
  and learnable neural population code.
\newblock {\em PNAS}, 108(23):9679--9684, 2011.

\bibitem{gantmacher:66}
F.~R. Gantmacher.
\newblock {\em the theory of matrices}.
\newblock AMS Chelsea Publishing, Providence, RI, 1998.

\bibitem{georgii:88}
Hans-Otto Georgii.
\newblock {\em Gibbs measures and phase transitions}.
\newblock De Gruyter Studies in Mathematics:9. Berlin; New York, 1988.

\bibitem{hammersley-clifford:71}
J.~M. Hammersley and P.~Clifford.
\newblock Markov fields on finite graphs and lattices.
\newblock {\em unpublished}, 1971.

\bibitem{harris-etal:02}
K.~D. Harris, D.~A. Henze, H.~Hirase, X.~Leinekugel, G.~Dragoi, A.~Czurko, ,
  and G.~Buzsaki.
\newblock Spike train dynamics predicts theta-related phase precession in
  hippocampal pyramidal cells.
\newblock {\em Nature}, 417(6890):738--741, 2002.

\bibitem{hastings:70}
W.K. Hastings.
\newblock Monte carlo sampling methods using markov chains and their
  applications.
\newblock {\em Biometrika}, 57:97--109, 1970.

\bibitem{higuchi-mezard:09}
S~Higuchi and M~Mezard.
\newblock Susceptibility propagation for constraint satisfaction problems.
\newblock {\em CoRR}, pages --1--1, 2009.

\bibitem{ikegay-etal:04}
Y.~Ikegaya, G.~Aaron, R.~Cossart, D.~Aronov, I.~Lampl, D.~Ferster, , and
  R.~Yuste.
\newblock Synfire chains and cortical songs: Temporal modules of cortical
  activity.
\newblock {\em Science}, 304(5670):559--564, 2004.

\bibitem{jaynes:57}
E.T. Jaynes.
\newblock Information theory and statistical mechanics.
\newblock {\em Phys. Rev.}, 106:620, 1957.

\bibitem{kappen-rodriguez:98}
H.J. Kappen and F.B. Rodriguez.
\newblock Boltzmann machine learning using mean field theory and linear
  response correction.
\newblock In {\em Advances in Neural Information Processing Systems}, pages
  280--286. The MIT Press, 1998.

\bibitem{keller:98}
G.~Keller.
\newblock {\em Equilibrium States in Ergodic Theory}.
\newblock Cambridge University Press, 1998.

\bibitem{kenet-etal:03}
T.~Kenet, D.~Bibitchkov, M.~Tsodyks, A.~Grinvald, , and A.~Arieli.
\newblock Spontaneously emerging cortical representations of visual attributes.
\newblock {\em Nature}, 425(6961):954--956, 2003.

\bibitem{lampl-etal:99}
I.~Lampl, I.~Reichova, and D.~Ferster.
\newblock Synchronous membrane potential fluctuations in neurons of the cat
  visual cortex.
\newblock {\em Neuron}, 22(2):361--374, 1999.

\bibitem{louie-wilson:01}
K.~Louie and M.~A. Wilson.
\newblock Temporally structured replay of awake hippocampal ensemble activity
  during rapid eye movement sleep.
\newblock {\em Neuron}, 29(1):145--156, 2001.

\bibitem{luczak-etal:09}
A.~Luczak, P.~Barth\'{o}, and K.~D Harris.
\newblock Spontaneous events outline the realm of possible sensory responses in
  neocortical populations.
\newblock {\em Neuron}, 62(3):413--425, 2009.

\bibitem{macke-cunningham-etal:11}
Jakob~H. Macke, John~P. Cunningham, Krishna~V. Shenoy, Lars Büsing, Byron~M.
  Yu, and Maneesh Sahani.
\newblock Empirical models of spiking in neural populations, 2010.

\bibitem{macke-etal:11}
J.H Macke, L~Busing, J.P Cunningham, B.M Yu, K.V Shenoy, and M~Sahani.
\newblock Empirical models of spiking in neural populations.
\newblock {\em NIPS}, 2001.

\bibitem{marre-boustani-etal:09}
O.~Marre, S.~El Boustani, Y.~Fr\'egnac, and A.~Destexhe.
\newblock Prediction of spatiotemporal patterns of neural activity from
  pairwise correlations.
\newblock {\em Phys. rev. Let.}, 102:138101, 2009.

\bibitem{mccullagh-nelder:89}
P.~McCullagh and J.~A. Nelder.
\newblock {\em Generalized linear models ({S}econd edition)}.
\newblock London: Chapman \& Hall, 1989.

\bibitem{mezard-mora:09}
M.~M\'ezard and T.~Mora.
\newblock Constraint satisfaction problems and neural networks: a statistical
  physics perspective.
\newblock {\em J. Physiol. Paris}, 103:107--113, 2009.

\bibitem{mokeichev-etal:07}
A.~Mokeichev, M.~Okun, O.~Barak, Y.~Katz, O.~Ben-Shahar, and I.~Lampl.
\newblock Stochastic emergence of repeating cortical motifs in spontaneous
  membrane potential fluctuations in vivo.
\newblock {\em Neuron}, 53(3):413--425, 2007.

\bibitem{nirenberg-victor-etal:07}
Sheila~H. Nirenberg and Jonathan~D. Victor.
\newblock Analyzing the activity of large populations of neurons: how tractable
  is the problem?
\newblock {\em Current Opinion in Neurobiology}, 17(4):397--400, 2007.

\bibitem{ohiorhenuan-mechler-etal:10}
Ifije~E. Ohiorhenuan, Ferenc Mechler, Keith~P. Purpura, Anita~M. Schmid, Qin
  Hu, and Jonathan~D. Victor.
\newblock Sparse coding and high-order correlations in fine-scale cortical
  networks.
\newblock {\em Nature}, 466(7306):617--621, 2010.

\bibitem{oram-etal:99}
M.~W. Oram, M.~C. Wiener, R.~Lestienne, and B.~J. Richmond.
\newblock Stochastic nature of precisely timed spike patterns in visual system
  neuronal responses.
\newblock {\em J Neurophysiol}, 81(6):3021--3033, 1999.

\bibitem{paninski-fellows-etal:04}
L.~Paninski, M.~Fellows, S.~Shoham, N.~Hatsopoulos, and J.~Donoghue.
\newblock {Superlinear population encoding of dynamic hand trajectory in
  primary motor cortex}.
\newblock {\em J. Neurosci.}, 24:8551--8561, 2004.

\bibitem{paninski:04}
Liam Paninski.
\newblock {Maximum likelihood estimation of cascade point-process neural
  encoding models}.
\newblock {\em Network: Comput. Neural Syst.}, 15(04):243--262, November 2004.

\bibitem{parry-pollicott:90}
W.~Parry and M.~Pollicott.
\newblock {\em Zeta functions and the periodic orbit structure of hyperbolic
  dynamics}, volume 187--188.
\newblock Asterisque, 1990.

\bibitem{pillow-shlens-etal:08}
J~W Pillow, J~Shlens, L~Paninski, A~Sher, A~M Litke, E~J Chichilnisky, and E~P
  Simoncelli.
\newblock Spatio-temporal correlations and visual signaling in a complete
  neuronal population.
\newblock {\em Nature}, 454(7206):995--999, Aug 2008.

\bibitem{pillow-ahmadian-etal:11b}
Jonathan~W. Pillow, Yashar Ahmadian, and Liam Paninski.
\newblock Model-based decoding, information estimation, and change-point
  detection techniques for multineuron spike trains.
\newblock {\em Neural Comput.}, 23(1):1--45, 2011.

\bibitem{pillow-paninski-etal:05}
J.W. Pillow, L.~Paninski, V.J. Uzzell, E.P. Simoncelli, and E.J. Chichilnisky.
\newblock Prediction and decoding of retinal ganglion cell responses with a
  probabilistic spiking model.
\newblock {\em Journal of Neuroscience}, 25(47):11003--11013, 2005.

\bibitem{pollicott-weiss:03}
Mark Pollicott and Howard Weiss.
\newblock Free energy as a dynamical invariant (or can you hear the shape of a
  potential?).
\newblock {\em Communications in Mathematical Physics}, 240:457--482, 2003.

\bibitem{puchalla-schneidman-etal:05}
J.~L. Puchalla, E.~Schneidman, R.~A. Harris, and M.~J. Berry.
\newblock Redundancy in the population code of the retina.
\newblock {\em Neuron}, 46(3):493--504, 2005.

\bibitem{roudi-nirenberg-etal:09}
Y.~Roudi, S.~Nirenberg, and P.E. Latham.
\newblock Pairwise maximum entropy models for studying large biological
  systems: when they can work and when they can't.
\newblock {\em PLOS Computational Biology}, 5(5), 2009.

\bibitem{roudi-hertz:11}
Yasser Roudi and John Hertz.
\newblock Mean field theory for non-equilibrium network reconstruction.
\newblock {\em Phys. Rev. Lett.}, 106(048702), 2011.

\bibitem{roudi-tyrcha-etal:09}
Yasser Roudi, Joanna Tyrcha, and John~A Hertz.
\newblock Ising model for neural data: Model quality and approximate methods
  for extracting functional connectivity.
\newblock {\em Physical Review E}, page 051915, 2009.

\bibitem{ruelle:78}
D.~Ruelle.
\newblock {\em Thermodynamic formalism}.
\newblock Addison-Wesley,Reading, Massachusetts, 1978.

\bibitem{schneidman-berry-etal:06}
E.~Schneidman, M.J. Berry, R.~Segev, and W.~Bialek.
\newblock Weak pairwise correlations imply strongly correlated network states
  in a neural population.
\newblock {\em Nature}, 440(7087):1007--1012, 2006.

\bibitem{seneta:06}
E.~Seneta.
\newblock {\em Non-negative Matrices and Markov Chains}.
\newblock Springer, 2006.

\bibitem{sessak-monasson:09}
V~Sessak and R~Monasson.
\newblock Small-correlation expansions for the inverse ising problem.
\newblock {\em Journal of Physics A}, 42(055001), 2009.

\bibitem{shadlen-newsome:98}
M.~Shadlen and W.~Newsome.
\newblock The variable discharge of cortical neurons: Implications for
  connectivity.
\newblock {\em J. Neurosci}, 18:3870--3896, 1998.

\bibitem{shenoy-matthew-etal:11}
Krishna~V. Shenoy, Matthew~T. Kaufman, Maneesh Sahani, and Mark~M. Churchland.
\newblock dynamical systems view of motor preparation: Implications for neural
  prosthetic system design.
\newblock {\em Progress in Brain Research: Enhancing Performance for Action and
  Perception}, 192, 2011.

\bibitem{sherrington-kirkpatrick:75}
D.~Sherrington and S.~Kirkpatrick.
\newblock Solvable model of a spin-glass.
\newblock {\em Physical Review Letters}, 35(26):1792+, December 1975.

\bibitem{shlens-field-etal:06}
J.~Shlens, G.D. Field, J.L. Gauthier, M.I. Grivich, D.~Petrusca, A.~Sher, A.M.
  Litke, and E.J. Chichilnisky.
\newblock The structure of multi-neuron firing patterns in primate retina.
\newblock {\em Journal of Neuroscience}, 26(32):8254, 2006.

\bibitem{shlens-field-etal:09}
Jonathon Shlens, Greg~D. Field, Jeffrey~L. Gauthier, Martin Greschner,
  Alexander Sher, Alan~M. Litke, and E.~J. Chichilnisky.
\newblock The structure of large-scale synchronized firing in primate retina.
\newblock {\em The Journal of Neuroscience}, 29(15):5022--5031, April 2009.

\bibitem{simoncelli-paninski-etal:04}
E.~P. Simoncelli, J.~P. Paninski, J.~Pillow, and O.~Schwartz.
\newblock {Characterization of Neural Responses with Stochastic Stimuli}.
\newblock {\em The cognitive neurosciences}, 2004.

\bibitem{singer-gray:95}
W.~Singer and C.~M. Gray.
\newblock Visual feature integration and the temporal correlation hypothesis.
\newblock {\em Annual Review of Neuroscience}, 18(1):555--586, 1995.

\bibitem{softky-Koch:93}
William~R. Softky and Christof Koch.
\newblock The highly irregular firing of cortical cells is inconsistent with
  temporal integration of random epsps.
\newblock {\em Journal of Neuroscience}, 13:334--350, 1993.

\bibitem{staude-gruen-etal:10}
B.~Staude, S.~Grun, and S.~Rotter.
\newblock Higher-order correlations in non-stationary parallel spike trains:
  statistical modeling and inference.
\newblock {\em Frontiers in computational neuroscience}, 4, 2010.

\bibitem{strong-koberle-etal:98}
S.P. Strong, R.~Koberle, R.R. de~Ruyter~van Steveninck, and W.~Bialek.
\newblock Entropy and information in neural spike trains.
\newblock {\em Phys. Rev. Let}, 80(1):197--200, 1998.

\bibitem{tanaka:98}
T.~Tanaka.
\newblock A theory of mean field approximation.
\newblock {\em ADVANCES IN NEURAL INFORMATION PROCESSING SYSTEMS}, 48:351--360,
  1998.

\bibitem{tang-jackson-etal:08}
Aonan Tang, David Jackson, Jon Hobbs, Wei Chen, Jodi~L. Smith, Hema Patel,
  Anita Prieto, Dumitru Petrusca, Matthew~I. Grivich, Alexander Sher, Pawel
  Hottowy, Wladyslaw Dabrowski, Alan~M. Litke, and John~M. Beggs.
\newblock A maximum entropy model applied to spatial and temporal correlations
  from cortical networks \textit{In Vitro}.
\newblock {\em The Journal of Neuroscience}, 28(2):505--518, January 2008.

\bibitem{theunissen-david-etal:01}
F.~E. Theunissen, S.~V. David, N.~C. Singh, A.~Hsu, W.~E. Vinje, and J.~L.
  Gallant.
\newblock {Estimating spatio-temporal receptive fields of auditory and visual
  neurons from their responses to natural stimuli}.
\newblock {\em Network}, 12(3):289--316, January 2001.

\bibitem{thouless-anderson-etal:77}
D.~J. Thouless, P.~W. Anderson, and R.~G. Palmer.
\newblock {Solution of solvable model of a spin glass}.
\newblock {\em Phil. Mag.}, 35:593--601, 1977.

\bibitem{tkacik-prentice-etal:10}
Gasper Tkacik, Jason~S. Prentice, Vijay Balasubramanian, and Elad Schneidman.
\newblock Optimal population coding by noisy spiking neurons.
\newblock {\em PNAS}, 107(32):14419--14424, August 2010.

\bibitem{tkacik-schneidman-etal:09}
Gasper Tkacik, Elad Schneidman, Michael~J. {Berry II}, and William Bialek.
\newblock Spin glass models for a network of real neurons.
\newblock {\em arXiv: 0912.5409v1}, 2009.

\bibitem{truccolo-eden-etal:05}
Wilson Truccolo, Uri~T. Eden, Matthew~R. Fellows, John~P. Donoghue, and
  Emery~N. Brown.
\newblock A point process framework for relating neural spiking activity to
  spiking history, neural ensemble and extrinsic covariate effects.
\newblock {\em J Neurophysiol}, 93:1074--1089, 2005.

\bibitem{tsodyks-etal:99}
M.~Tsodyks, T.~Kenet, A.~Grinvald, , and A.~Arieli.
\newblock Linking spontaneous activity of single cortical neurons and the
  underlying functional architecture.
\newblock {\em Science}, 286(5446):1943--1946, 1999.

\bibitem{vaadia-etal:95}
E.~Vaadia, I.Haalman, M.~Abeles, H.~Bergman, Y.Prut, H.~Slovin, and A.~Aertsen.
\newblock Dynamics of neuronal interactions in monkey cortex in relation to
  behavioural events.
\newblock {\em Nature}, 373(6514):515--518, 1995.

\bibitem{vasquez-marre-etal:12}
Juan~Carlos Vasquez, Olivier Marre, Adrian~G Palacios, Michael~J Berry, and
  Bruno Cessac.
\newblock Gibbs distribution analysis of temporal correlation structure on
  multicell spike trains from retina ganglion cells.
\newblock {\em J. Physiol. Paris}, 2012.
\newblock in press.

\bibitem{welling-teh:11}
Max Welling and Yee~Whye Teh.
\newblock Approximate inference in boltzmann machines.
\newblock {\em Artificial Intelligence}, 143(1):19 -- 50, 2003.

\bibitem{yu-huang-etal:08}
Shan Yu, Debin Huang, Wolf Singer, and Danko Nikolic.
\newblock A small world of neuronal synchrony.
\newblock {\em Cereb. Cortex}, 2008.

\end{thebibliography}
